\documentclass[%
superscriptaddress,
floats,
floatfix,
twocolumn,
nofootinbib,
amsmath,amssymb,
aps
]{revtex4-1}

\usepackage{array}
\usepackage{dcolumn}% Align table columns on decimal point
\usepackage{bm}% bold math
\usepackage[colorlinks, pdfborder={0 0 0}, plainpages=false]{hyperref}
\usepackage{graphicx}
\usepackage{xspace}
\usepackage[usenames,dvipsnames]{color}
\usepackage{amssymb}
\usepackage[normalem]{ulem} %for \sout
\usepackage{lineno}
\usepackage{letltxmacro}
\usepackage{amsfonts}
\usepackage{amsmath}
\usepackage{amssymb}
\usepackage{bm}
\usepackage{dcolumn}
\usepackage{epsfig}
\usepackage{graphicx}
\usepackage{graphics}
\usepackage[latin1]{inputenc}
\usepackage{latexsym}
\usepackage{rotating}
\usepackage{hyperref}
\usepackage[usenames]{color}
\usepackage{float}
\usepackage{xspace} 
\usepackage{mathrsfs}
\usepackage{multirow}
\usepackage{booktabs}
\usepackage{soul}
\usepackage{subcaption}

\usepackage{colortbl}
\usepackage{etoolbox}
\usepackage[table]{xcolor}

\def\apj{{\it Astrophys. J.}}

\def\mnras{{\it MNRAS~}}

\def\aj{{\it The Astronomical Journal~}}
\def\jmlr{{\it Journal of Machine Learning Research}}

\def\lesssim{\mathrel{\hbox{\rlap{\hbox{\lower4pt\hbox{$\sim$}}}\hbox{$<$}}}}
\def\gtrsim{\mathrel{\hbox{\rlap{\hbox{\lower4pt\hbox{$\sim$}}}\hbox{$>$}}}}
\def\alt{\mathrel{\hbox{\rlap{\hbox{\lower4pt\hbox{$\sim$}}}\hbox{$<$}}}}
\def\agt{\mathrel{\hbox{\rlap{\hbox{\lower4pt\hbox{$\sim$}}}\hbox{$>$}}}}
\def\apjs{{\it ApJS}}

\usepackage{mathtools}

\newenvironment{cititemize2}
{\begin{list}{$\bullet$}
        {\setlength{\topsep}{0pt}
         \setlength{\itemsep}{0pt}
         \setlength{\parsep}{0.25\parsep}
         \settowidth{\labelwidth}{$\bullet$}
         \setlength{\leftmargin}{1em}
}
}
{\end{list}}

\def\gta{\ifmmode {\mathbin{\lower 3pt\hbox   %> or of order
    {$\,\rlap{\raise 5pt\hbox{$\char'076$}}\mathchar"7218\,$}}}
    \else {${\mathbin{\lower 3pt\hbox
    {$\rlap{\raise 5pt\hbox{$\char'076$}}\mathchar"7218\,$}}}
    $}\fi}
\def\lta{\ifmmode {\,\mathbin{\lower 3pt\hbox   %< or of order
    {$\,\rlap{\raise 5pt\hbox{$\char'074$}}\mathchar"7218\,$}}}
    \else {${\mathbin{\lower 3pt\hbox
    {$\rlap{\raise 5pt\hbox{$\char'074$}}\mathchar"7218\,$}}}
    $}\fi}

\newcommand{\beq}{\begin{equation}}
\newcommand{\eeq}{\end{equation}}
\newcommand{\bea}{\begin{eqnarray}}
\newcommand{\eea}{\end{eqnarray}}

\definecolor{darkperiwinkle}{RGB}{102, 102, 128}

\newcommand{\NCSA}{\affiliation{NCSA, University of Illinois at Urbana-Champaign, Urbana, Illinois 61801, USA}}
\newcommand{\ANCSA}{\affiliation{Department of Astronomy, University of Illinois at Urbana-Champaign, Urbana, Illinois 61801, USA}}
\newcommand{\PNCSA}{\affiliation{Department of Physics, University of Illinois at Urbana-Champaign, Urbana, Illinois 61801, USA}}
\newcommand{\ALCF}{\affiliation{Argonne National Laboratory, Leadership Computing Facility, Lemont, Illinois 60439, USA}}

\definecolor{light-gray}{gray}{0.9}

\begin{document}

\title{Deep Learning at Scale\\ for the Construction of Galaxy Catalogs\\ in the Dark Energy Survey}

\author{Asad Khan}\NCSA\PNCSA
\author{E. A. Huerta}\NCSA\ANCSA
\author{Sibo Wang}\NCSA
\author{Robert Gruendl}\NCSA\ANCSA
\author{Elise Jennings}\ALCF
\author{Huihuo Zheng}\ALCF

\date{\today}

\begin{abstract}
\noindent The scale of ongoing and future electromagnetic surveys pose formidable challenges to classify astronomical objects. Pioneering efforts on this front include citizen science campaigns adopted by the Sloan Digital Sky Survey (SDSS). SDSS datasets have been recently used to train neural network models to classify galaxies in the Dark Energy Survey (DES) that overlap the footprint of both surveys. Herein, we demonstrate that knowledge from deep learning algorithms, pre-trained with real-object images, can be transferred to classify galaxies that overlap both SDSS and DES surveys, achieving state-of-the-art accuracy \(\gtrsim99.6\%\). We demonstrate that this process can be completed within just eight minutes using distributed training. While this represents a significant step towards the classification of DES galaxies that overlap previous surveys, we need to initiate the characterization of unlabelled DES galaxies in new regions of parameter space. To accelerate this program, we use our neural network classifier to label over ten thousand unlabelled DES galaxies, which do not overlap previous surveys. Furthermore, we use our neural network model as a feature extractor for unsupervised clustering and find that unlabeled DES images can be grouped together in two distinct galaxy classes based on their morphology, which provides a heuristic check that the learning is successfully transferred to the classification of unlabelled DES images. We conclude by showing that these newly labeled datasets can be combined with unsupervised recursive training to create large-scale DES galaxy catalogs in preparation for the Large Synoptic Survey Telescope era.
\end{abstract}

\pacs{Valid PACS appear here}% PACS, the Physics and Astronomy
                             % Classification Scheme.
%\keywords{Suggested keywords}%Use showkeys class option if keyword
                              %display desired
\maketitle

\noindent \textbf{Keywords}: Deep Learning, Convolutional Neural Networks, Sloan Digital Sky Survey, Dark Energy Survey, Large Synoptic Survey Telescope, Galaxy Catalogs, Unsupervised Learning, Data Clustering
%%%%%%%%%%%%%%%%%%%%%%%%%%%%%%%%%%%%%%%%%%%%%
%%%%%%%%%%%%%%%%%%%%%%%%%%%%%%%%%%%%%%%%%%%%%
\section{Introduction}
\label{intro}
%what is the goal of EM surveys
Electromagnetic surveys provide key insights into the large scale structure of
the Universe, its geometry and evolution in cosmic time. As the depth and scale of these 
surveys continue to increase in years to come, they will push back the frontiers of our 
understanding of dark matter and dark energy~\cite{Riess:1998AJR,Perl:1999ApJP,Tonry:2003ApJT,Knop:2003ApJK}. 

%what people did in the past
The classification of astrophysical objects has been pursued in the past using a diverse set of tools. For instance, galaxies have been classified using their photometric properties, achieving classification accuracies \(\sim85\%\)~\cite{photo_class}. Other methods to classify galaxies according to their morphology have taken into account their physical properties across multiple wavelengths. For instance, the method introduced in~\cite{class_sdss}, considered a sample of galaxies from the Sloan Digital Sky Survey (SDSS)~\cite{SDSS:2011AJ}, using the five SDSS filters \((u,\,g,\,r,\,i,\,z)\), and then used a combination of shapelet decomposition and Principal Components Analysis (PCA). Other methods for galaxy classification include Concentration-Asymmetry-Smoothness (CAS)~\cite{CAS_2003}, and machine learning, including artificial neural networks and PCAs~\cite{ann_galaxy,aan_II,banerji_2010}. 

%why are we doing this
In recent years, citizen science campaigns have played a key role to classify thousands of celestial objects in astronomical surveys. SDSS  is an archetypical example of a successful approach to classify hundreds of thousands of galaxies. As electromagnetic surveys continue to increase their depth and coverage, campaigns of this nature may lack scalability. For instance, within six years of operation, the Dark Energy Survey (DES)~\cite{DES:2016MNRASOv} observed over three hundred million galaxies, a number that will be surpassed by the observing capabilities of the Large Synoptic Survey Telescope (LSST)~\cite{LSST:2012L}. In brief, there is a pressing need to explore new approaches to maximize the science throughput of next-generation electromagnetic surveys. A promising paradigm is the convergence of deep learning and large scale computing to address the imminent increase in data volume, complexity, and latency of observations of LSST-type surveys, the theme of this paper. 

%what can be done
An innovative idea to accomplish this prospect consists of leveraging what SDSS has already done, and try to use it as seed information to classify objects in DES data. Such idea has been explored in~\cite{Dom:2018D}, where SDSS galaxies that overlap the DES footprint were used to train neural network models to classify DES galaxies that were also observed by SDSS, reporting classification accuracies \(\sim 95\%\)~\cite{Dom:2018D}. 

%what we do
While the aforementioned approach provides a way to classify DES galaxies that overlap previous surveys, key issues remain: (i) deep learning algorithms for image classification have been trained with \textit{hundreds of millions} of images to achieve state-of-the-art classification accuracy~\cite{chollet:2016}. If one attempts to train a neural network model from the ground up using just a few tens of thousands of SDSS galaxies, then the fully trained neural network model may not achieve state-of-the-art classification accuracy, or exhibit overfitting~\cite{shengeorge:PhysRevD97}; (ii) while training a neural network model with SDSS galaxies, and then applying it to classify DES galaxies that overlap the footprint of both SDSS and DES is an important validation study for the applicability of deep learning for classification analyses, we also need to demonstrate the applicability of this approach for DES galaxies that have not yet been observed in previous surveys. This can only be accomplished once more DES galaxies are labeled; (iii) newly labelled DES galaxies, that do not overlap previous surveys, can be used as training datasets to enhance the classification accuracy of deep learning algorithms. One can easily realize that this approach will lead to the creation of TB-size training datasets. In this scenario, it will be essential to design distributed algorithms to reduce the training stage at a minimum, while retaining state-of-the-art classification accuracy.

In this article we describe an approach to address the aforementioned challenges by bringing together several deep learning methods in an innovative manner. Key highlights of this study include:

\begin{cititemize2}
\item We transfer knowledge from the state-of-the-art neural network model for image classification, \texttt{Xception}~\cite{chollet:2016}, trained with the \texttt{ImageNet} dataset~\cite{DengCVPR2009}, to classify SDSS galaxy images, achieving state-of-the-art accuracies \(99.8\%\). Note that transfer learning between similar datasets, such as SDSS and DES, has been traditionally used in the computer science literature~\cite{dtl}. In stark contrast, we use a pre-trained model for  
real-world object recognition, and then transfer its knowledge to classify SDSS and DES galaxies. To the best of our knowledge this is the first application of deep transfer learning for galaxy classification\footnote{While this paper was under review, a study on SDSS galaxy classification was presented in which disparate datasets for transfer learning are used~\cite{barchi_2019}.}.
\item To streamline and accelerate this method, we introduce the first application of deep transfer learning and distributed training in cosmology, reducing the training stage of the \texttt{Xception} model with galaxy image datasets from five hours to just eight minutes, using 64 K80 GPUs in the Cooley supercomputer.
\item We show that our neural network model trained by transfer learning achieves state-of-the-art accuracy, \(99.6\%\), to classify DES galaxies that overlap the footprint of the SDSS survey. 
\item We use our neural network classifier to label over ten thousand unlabeled DES galaxies that have not been observed in previous surveys. We then turn our neural network model into a feature extractor to show that these unlabeled datasets can be clustered according to their morphology, forming two distinct datasets.
\item Finally, we use the newly labelled DES images and do unsupervised recursive training to retrain our deep transfer learning model, boosting its accuracy to 
classify unlabeled DES galaxies in bulk in new regions of parameter space.
\end{cititemize2}

The combination of all the aforementioned deep learning methods lays the foundations to exploit deep transfer learning at scale, data clustering and recursive training to produce large-scale galaxy catalogs in the LSST era~\cite{LSST:2012L}.

This paper is organized as follows. Section~\ref{method} presents the approach followed to curate the datasets 
and deep learning algorithms designed and trained for our analyses. In section~\ref{Clustering}, we demonstrate 
the applicability of our methods to classify galaxies in SDSS, galaxies that overlap SDSS and DES, and finally, the 
applicability of our approach to correctly classify thousands of unlabelled DES galaxies. 
Finally, section~\ref{end} summarizes our findings and future directions of work.

%%%%%%%%%%%%%%%%%%%%%%%%%%%%%%%%%%%%%%%%%%%%%
%%%%%%%%%%%%%%%%%%%%%%%%%%%%%%%%%%%%%%%%%%%%%
\section{Methods}
\label{method}

In this section we describe the SDSS and DES datasets we have curated for our studies, the neural network model we use for deep transfer learning, and how we can use unsupervised recursive training to create galaxy catalogs at scale.

\subsection{Data Curation for SDSS and DES}
\label{dc}

We use a subset of SDSS Data Release (DR) 7 images for which we have high confidence classifications through the Galaxy Zoo project, i.e., we only choose galaxies with debiased probability greater than 0.985 for combined spirals, and 0.926 for ellipticals, respectively, as shown in Table 2 of ~\cite{Lintott:2011MNRAS}. We choose these cutoff thresholds to ensure that; (i) the galaxies used for training the neural network have robust and accurate classifications; and (ii) the representation of both classes in the training and test datasets are balanced. We then divide these images into three separate datasets for training, validation and testing. The validation set is used to monitor the accuracy and loss when training and fine-tuning our deep neural network, and hence serves to optimize hyperparameters, such as learning rate and number of epochs, for training.

Two test sets are carefully constructed so that the images in each set lie in both the SDSS and DES footprints. The first test set consists of images with Galazy Zoo classification confidence similar to that of the training set, i.e., a high probability cut-off is introduced. This test set is hence labelled High Probability (HP) Test Set, and there are two versions, one for each survey, i.e., HP SDSS and HP DES.  Just as in the training set, the images for SDSS are obtained from (DR) 7 and the corresponding images for DES are obtained from the DES DR1 data release. Furthermore, a second test set is created without introducing any probability thresholds on Galaxy Zoo classification confidence. This set consists of almost all galaxies lying in both the SDSS and DES footprints, and 

%%%%%%%%%%%%%%%%%%%% DATA SUMMARY %%%%%%%%%%%%%%%%%%%%%%%%%
\begin{table}[h!]
		\footnotesize
		\begin{center}
                        \setlength{\tabcolsep}{19pt} % default is apparently 6pt
			\begin{tabular}{lc c c|}
				\hline 
				Dataset & Spirals & Ellipticals \\ 
				\hline
				Training set &  18,352 & 18,268 \\
				 \hline
				HP SDSS Test Set  & 516 & 550 \\
				HP DES\,\,\,   Test Set &  516& 550 \\
				\hline
				FO SDSS Test Set  & 6,677 & 5,904 \\
				FO DES\,\,\,   Test Set & 6,677& 5,904 \\
				\hline
			\end{tabular}
		\end{center}
	\caption{Summary of the SDSS and DES datasets used for training and testing.}
	\label{table:summary}
	\end{table}
	\normalsize
%%%%%%%%%%%%%%%%%%%% Prob Dist of each Data set %%%%%%%%%%%%%%%%%%%%%%%

\begin{figure}[!h]
\centerline{
\includegraphics[width=0.5\textwidth]{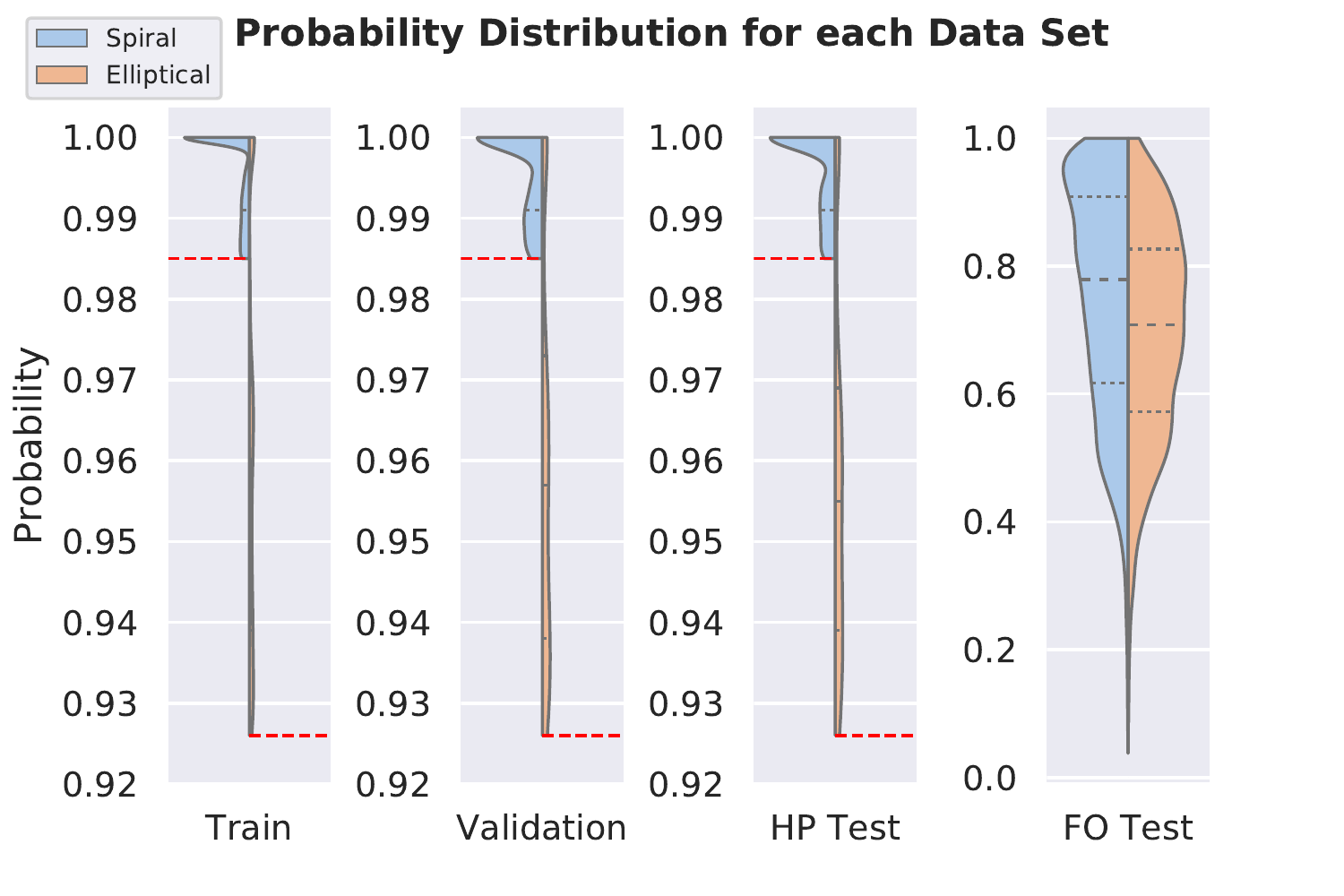}
}
\caption{Violin Plots of Galaxy Zoo Probability Distributions for galaxies in each dataset. Probability threshold cutoffs, 98.5\% for spiral and 92.6\% for elliptical, are shown as red dashed lines. These cutoffs have been selected to ensure that the datasets of both galaxy types are balanced.}
\label{fig:Prob_Dist}
\end{figure}

hence is labelled Full Overlap (FO) Test Set. Again there are two versions, FO SDSS and FO DES. The motivation behind creating this second test set is that the galaxy profiles in the unlabelled DES dataset will more closely match those in FO test sets. Hence FO test set serves as a good evaluation metric of the performance of our neural net on the ultimate task of classifying all unlabelled galaxies in the DES catalogue.

\begin{figure*}[!h]
\centerline{
\includegraphics[width=0.35\linewidth, height=0.25\linewidth]{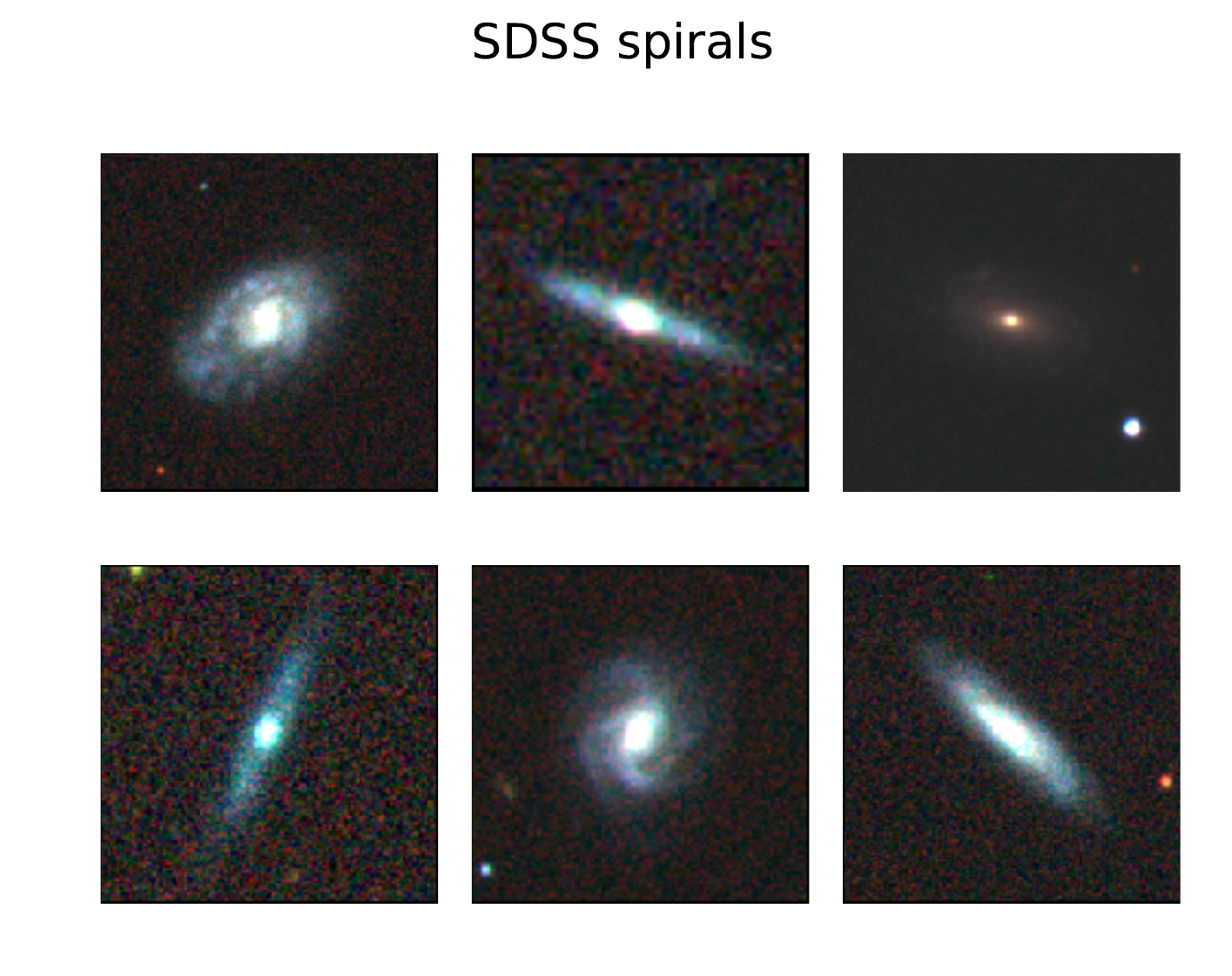}\hspace{-0.6em}%
\includegraphics[width=0.35\linewidth, height=0.25\linewidth]{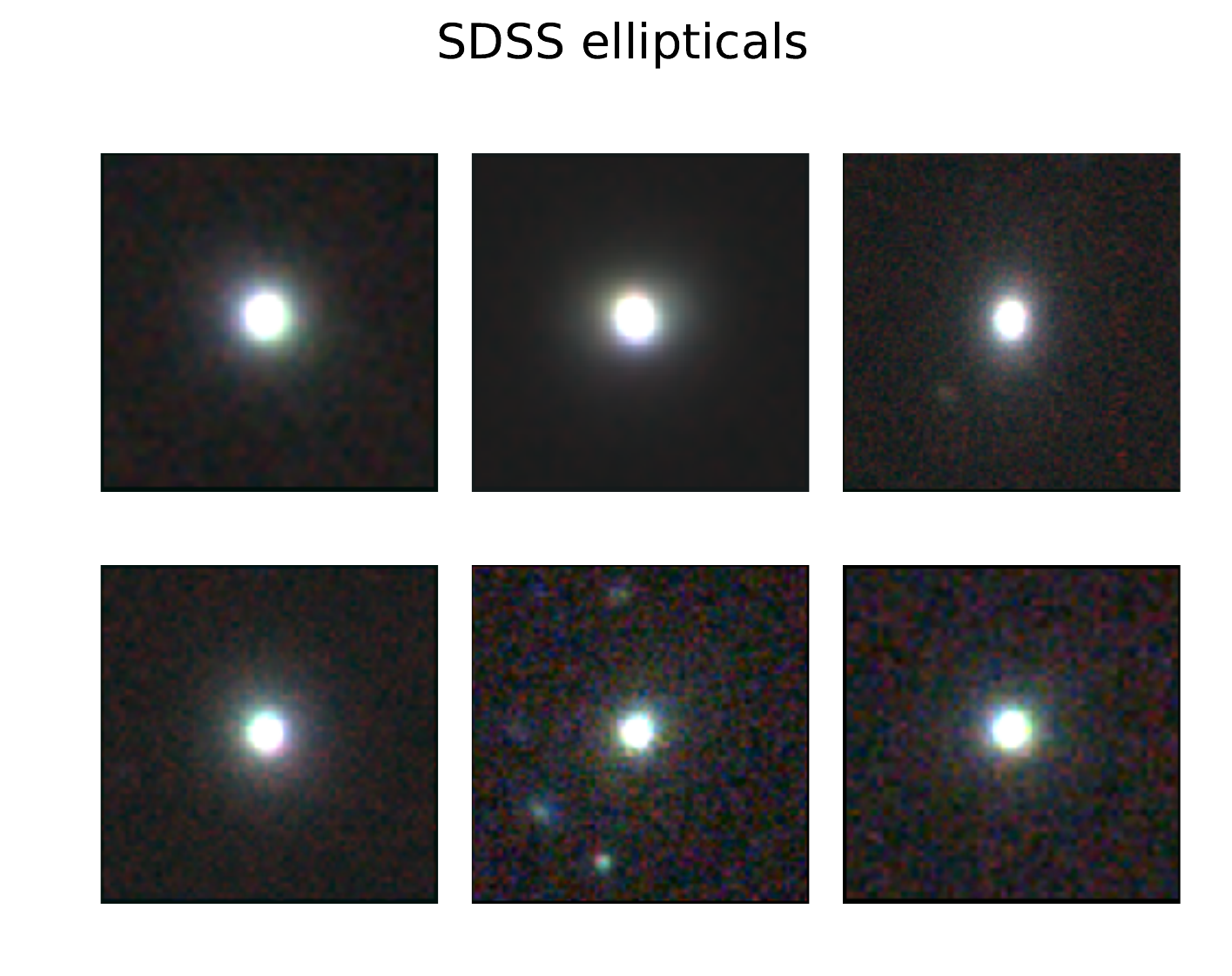}
}
\centerline{
\includegraphics[width=0.35\linewidth, height=0.25\linewidth]{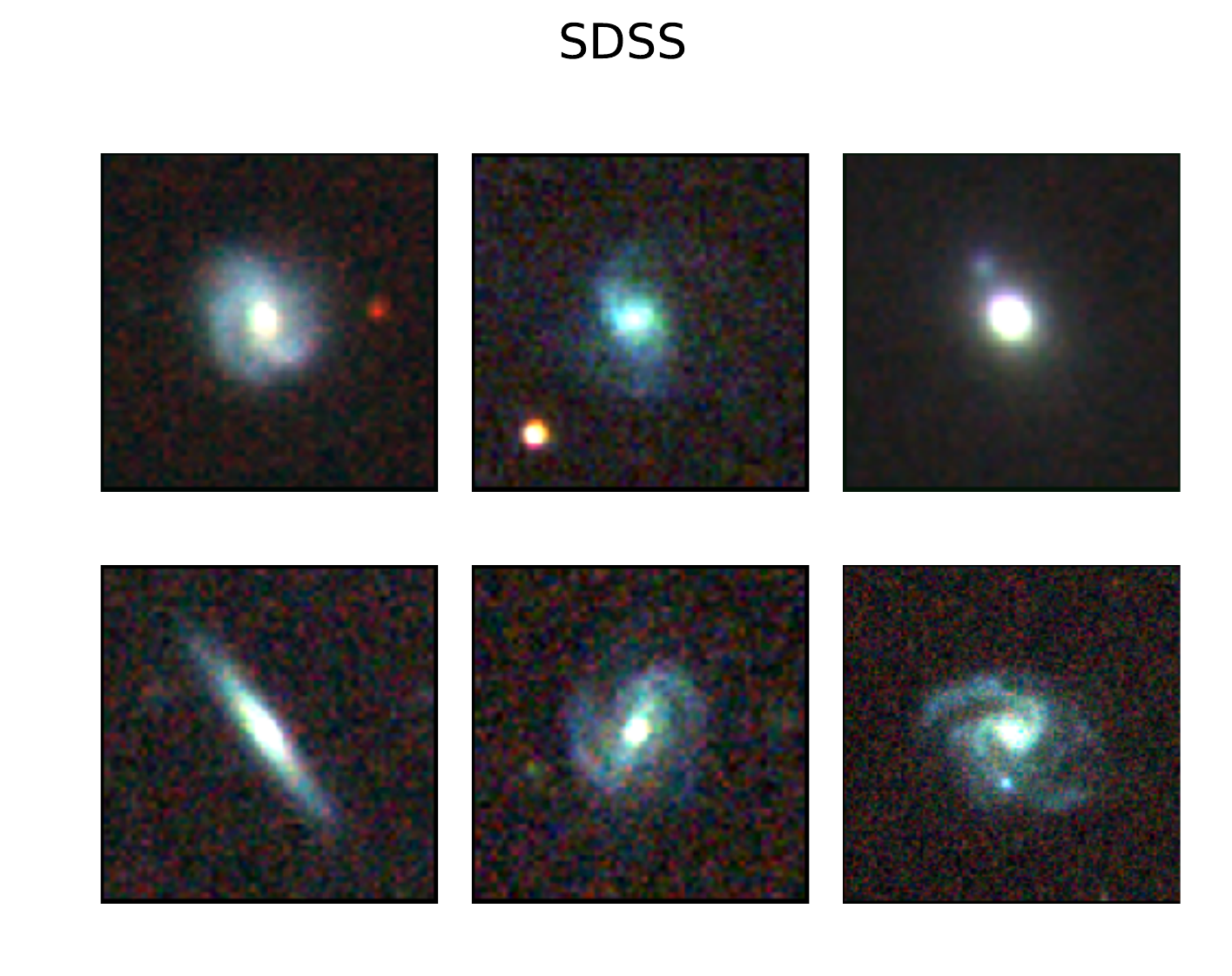}\hspace{-0.6em}%
\includegraphics[width=0.35\linewidth, height=0.25\linewidth]{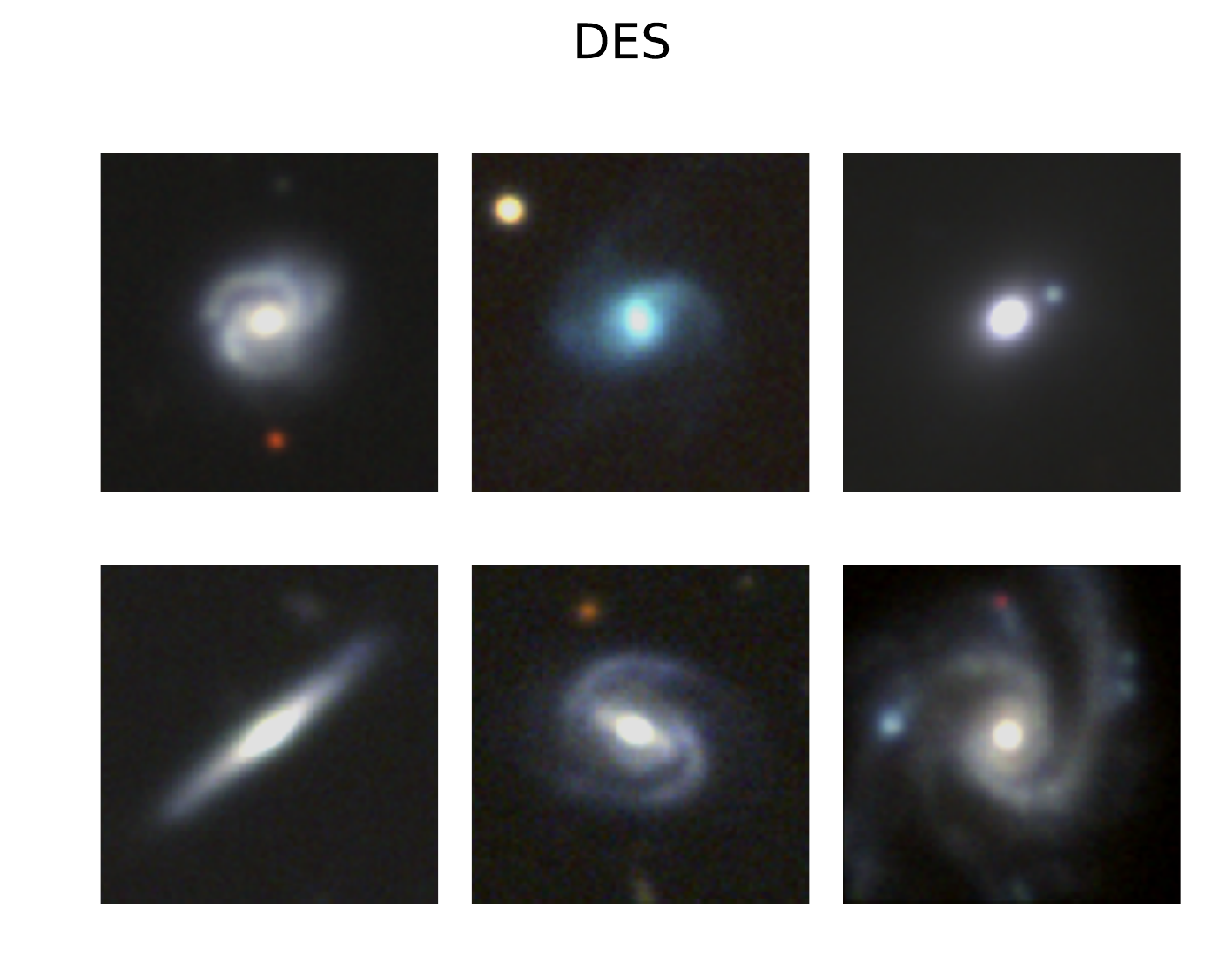}
}
\caption{Top panels: labelled images of the SDSS training set. Bottom panels: sample of galaxies from SDSS-DR7 and the corresponding crossmatched galaxies from DES DR1.}
\label{fig:images_SDSS}
\end{figure*}

The properties of these datasets are summarized in Table~\ref{table:summary}, while their probability distributions are presented in Fig.~\ref{fig:Prob_Dist}. A sample of the training SDSS dataset, and the HP Test set images are presented in the top and bottom panels of Fig.~\ref{fig:images_SDSS}, respectively.

\noindent {\bf SDSS Dataset} We used the de-biased probabilities for elliptical and combined spiral classes described in Table 2 of~\cite{Lintott:2011MNRAS} to create labels for the two classes of our training and test sets. After selecting the OBJIDs from Table 2 based on the probability thresholds of 0.985 and 0.926 for spirals and ellipticals respectively, we submit SQL queries to the SDSS Skyserver~\cite{SDSSServer} to obtain g, r and i-band images and metadata from the PhotoObj table. Thereafter, each galaxy is `cut-out' from the downloaded telescope fits files for each band, and then the bands are stacked together to create a color image. 

We developed the scripts to download and preprocess data as open source 
\texttt{Python} software stacks~\footnote{The code is publicly available in a github repository at \url{https://github.com/khanx169/DL_DES}}. To facilitate and streamline these tasks at scale, 
we incorporated Message Passing Interface (MPI)~\cite{MPIBook:1999} to exploit multiple nodes on
supercomputers for a fast parallel computation. In our case, the data extraction and 
curation was done using the Blue Waters Supercomputer~\cite{Kramer2015}.

\noindent {\bf DES Dataset} The same steps are repeated to first select the DES DR1 metadata and images from the NCSA DESaccess web~\cite{DESServer}, and then to cut-out, preprocess and stack the filters together to create RGB color images. Additionally, the Astropy package \texttt{match\_to\_catalog\_sky} is used to crossmatch DES and SDSS catalogues to within 1 arcsec. Finally we pick a random sample of \(\sim10,000\) bright DES galaxies to quantify the classification and clustering performance of our neural network model.

\subsection{Deep Learning: Model and Methodology}
\label{DeepLearning}

\begin{figure*}[!h]
\centerline{
\includegraphics[width=0.8\linewidth, height=0.3\linewidth]{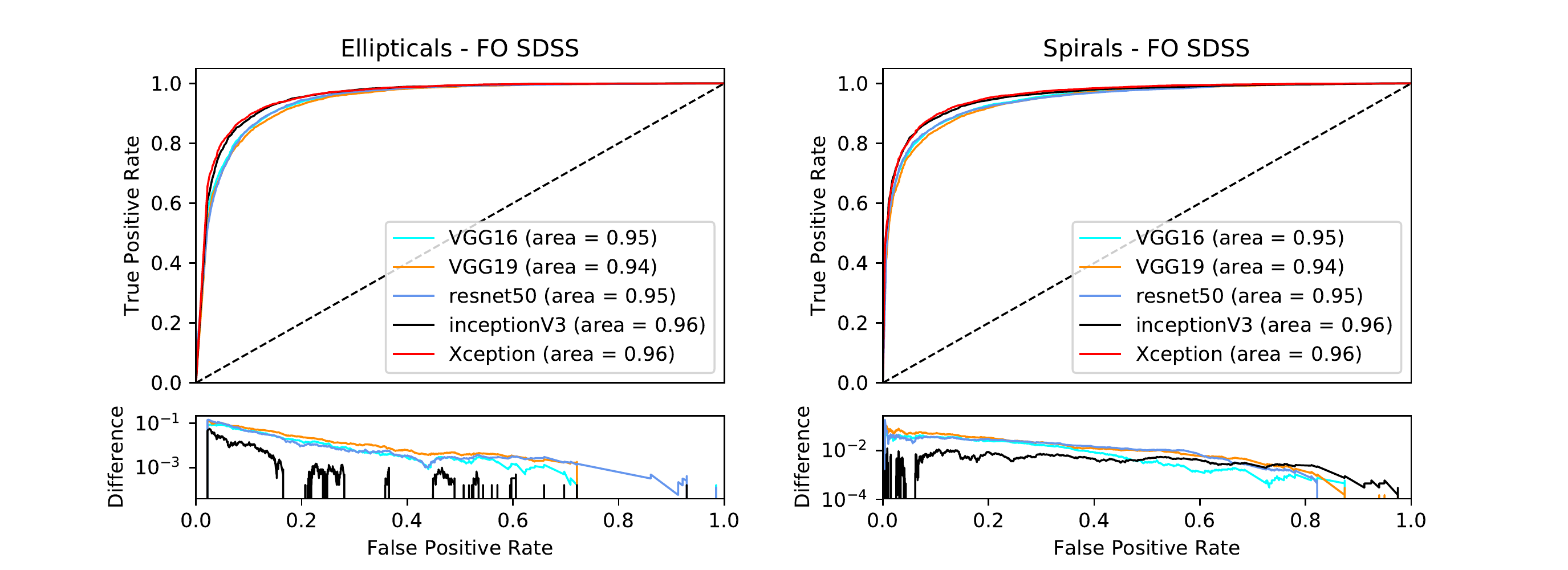}
}
\centerline{
\includegraphics[width=0.8\linewidth, height=0.3\linewidth]{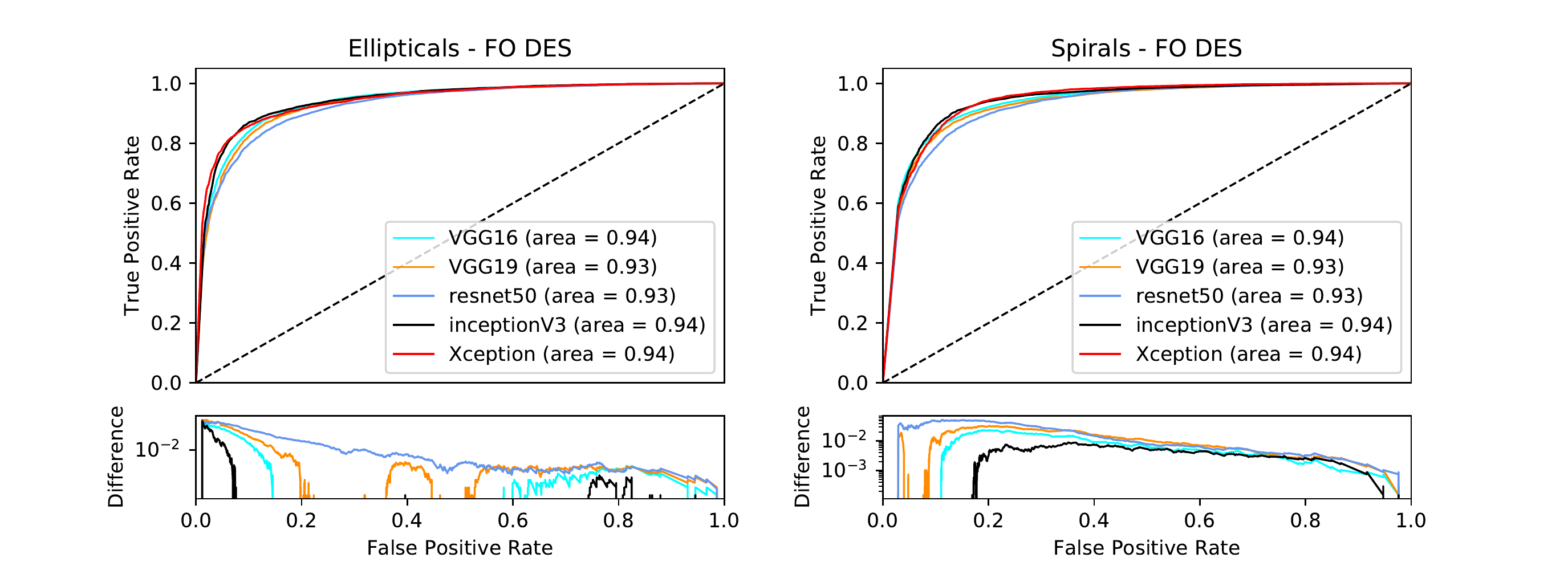}
}
\caption{Performance of several different fine-tuned architectures pre-trained on \texttt{ImageNet}. Top panels: Receiver Operating Characteristic (ROC) for galaxies in the FO SDSS test set. Bottom panels: ROC  for galaxies in the FO DES test set. The smaller insets show the difference in True Positive Rate between Xception and each of the other four models on a log scale.}
\label{fig:diff_models}
\end{figure*}

We use open source software stacks for our studies. The deep learning APIs used are \texttt{Keras}~\cite{Keras} and \texttt{Tensorflow}~\cite{TF}. For the classification problem we do transfer learning starting with the \texttt{Xception} model~\cite{chollet:2016}, which has been pre-trained with the \texttt{ImageNet}~\cite{ImNet:2014} dataset. We choose this neural network model because it outperforms many other state-of-the-art neural network models, including \texttt{Inception-v3}~\cite{Inception:2015}, \texttt{ResNet-152}~\cite{resnet:2015} and \texttt{VGG16}~\cite{vgg16:2014} on \texttt{ImageNet} validation dataset, and it has been suggested that better \texttt{ImageNet} architectures are capable of learning better transferable representations ~\cite{1805.08974}. More importantly, we carried out several experiments and found that \texttt{Xception} exhibits either as good or nominally better performance on our validation and testing galaxy datasets compared to many other state of the art architectures (see Fig.~\ref{fig:diff_models}).  

\noindent Bearing in mind that the \texttt{Xception} model~\cite{chollet:2016} was originally trained on the \texttt{ImageNet}~\cite{ImNet:2014} dataset, with images resized to
\(299\times299\times3\), we have followed best practices of neural network training~\cite{conimag}, and have resized all the galaxy sub-images to be \(299\times299\)
pixels, and then 
stacked the three filters together to create a color image of size 
\(299\times299\times3\). Finally, these sub-images are mean subtracted and normalized to convert the pixel values to -1 to 1 range centered 
around 0. These curated datasets serve as the input tensor into our deep neural network model.

For training, we first extract the feature maps from the second last layer of the pretrained model for a single epoch and feed them into a few custom defined fully connected layers added at the end of the pre-trained model (see Figure~\ref{fig:model} in~\ref{ap1}). Then we progressively unfreeze the earlier layers of the whole network and fine tune their weights for a few epochs of training.

%%%%%%%%%%%%%%%%%%%% Sample of Training Images %%%%%%%%%%%%%%%%%%%%%%%%%
%%%%%%%%%%%%%%%%%%%%%%%%%%%%%%%%%%%%%%%%%%%%%

%%%%%%%%%%%%%%%%%%%% Performance of various models %%%%%%%%%%%%%%%%%%%%%%%%%
%%%%%%%%%%%%%%%%%%%%%%%%%%%%%%%%%%%%%%%%%%%%%

The rationale behind this approach is that the earlier layers of a trained network are very versatile filters able to pick up simple abstract features like lines and edges relevant to any image detection or classification problem. However, deeper into the network, the weights of the layers become less interpretable and more specific to the given problem at hand~\cite{Zeiler2014VisualizingAU}. Hence, by training the last layers first and then progressively fine tuning the earlier layers we make sure that the useful weights learned on millions of \texttt{ImageNet}~\cite{DengCVPR2009} images are not destroyed while the neural network learns and adapts to the galaxy classification problem~\cite{Yosinski:2014:TFD:2969033.2969197}. Deep transfer learning has been explored in physics and astronomy classification problems, including noise anomaly classification in gravitational wave data~\cite{shengeorge:PhysRevD97}, galaxy merger classification~\cite{dtl_galaxy_merger}, and galaxy classification~\cite{barchi_2019,Dom:2018D}.

\noindent \textbf{Single-GPU Training} 
We train the network using Tesla P100 GPUs on XSEDE (Bridges)~\cite{BridgeX}. The training process for the dataset of 36500 images is completed within 5 hours. We use categorical cross entropy as the loss function together with ADAM optimizer~\cite{ADAM}. To avoid over-fitting, we monitor both training and validation losses, add a dropout rate of 70\% between our fully connected layers, and also use early-stopping, i.e., we stop training once validation loss stops decreasing. Additionally we use the learning rate scheduler, i.e., we reduce the learning rate when training loss stops decreasing to do a more fine-grained search of the loss function's minima, and data augmentation. For data augmentation we use random flips, rotations, zooms and shifts as shown in Figure~\ref{fig:data_augmentation} in~\ref{ap2}. After training, all the weights are frozen and saved, and inference on about 10,000 test images is completed within 10 minutes using a single Tesla P100 GPU.

\begin{figure}[h!]
\centerline{
\includegraphics[width=0.9\linewidth]{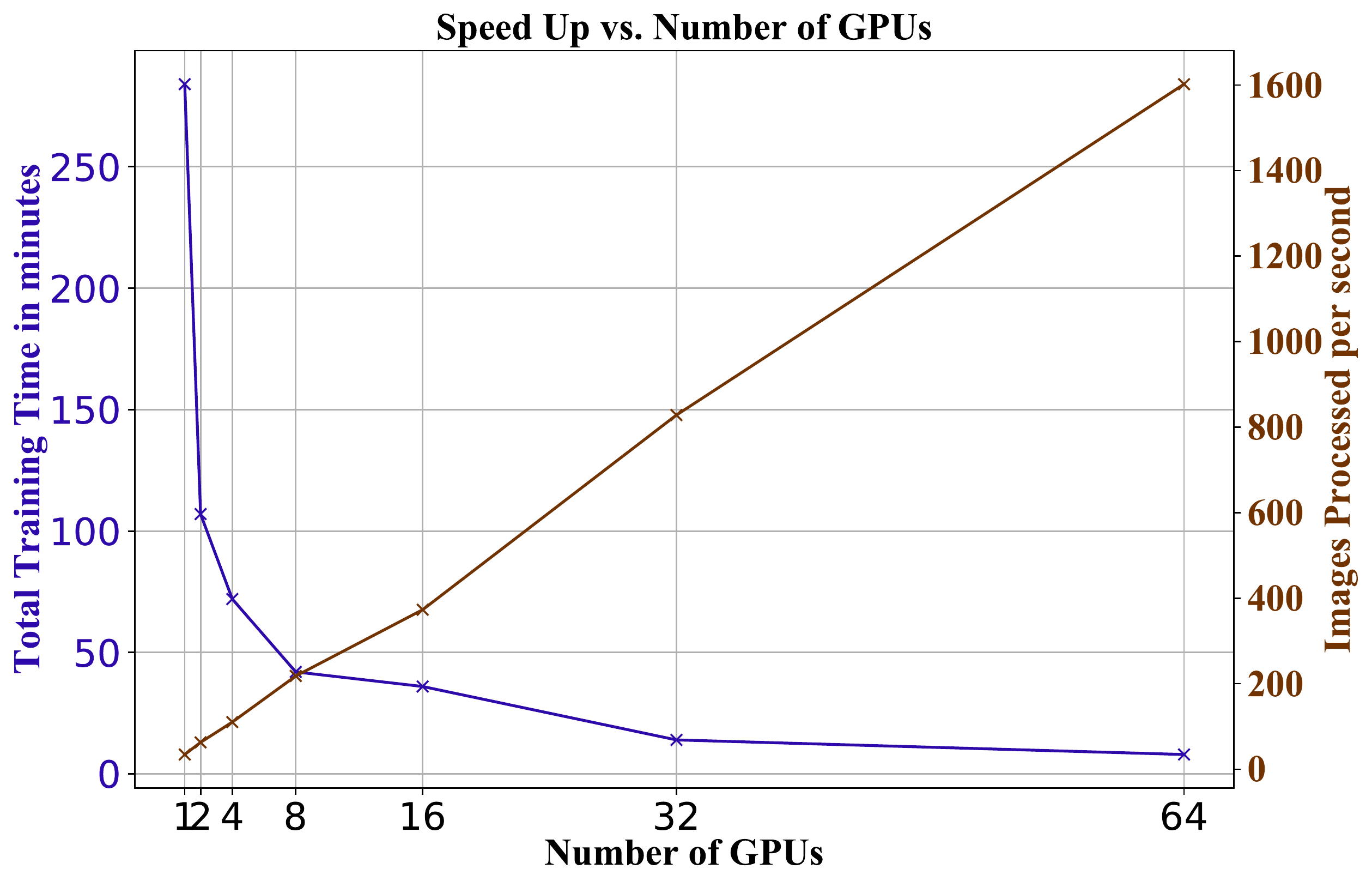}
} 
\caption{Speed up in training using up to 64 K80 GPUs for the \texttt{Xception} model.}
\label{fig:cooley}
\end{figure}

\noindent \textbf{Distributed Training}
\noindent Figure~\ref{fig:cooley} shows the parallel training performance for the \texttt{Xception} model using up to 64 K80 GPUs (see Table~\ref{table:summary-scaling-results} for a detailed breakdown of these results). The code was distributed across multiple GPUs using the \texttt{Horovod} distributed framework for \texttt{Keras}~\cite{sergeev2018horovod}.\footnote{These results were obtained on the Intel
Haswell and NVIDIA K80 based supercomputer, Cooley, at Argonne Leadership Computing Facility using a data parallelization scheme through Horovod.} 
We find that distributing the workload decreases the time per epoch linearly, and significantly reduces the training of 36,620 images from $\sim 5$ hours using a single GPU to 8m using 64 GPUs, with similar accuracy.~\footnote{ The base learning rate was 0.0001 and was scaled by the number of GPUs, $N$, following \cite{DBLP:journals/corr/GoyalDGNWKTJH17} while keeping the mini-batch size the same on each worker. In addition we used a technique of ``warmup'' epochs where we set
the learning rate to be the base learning rate and increase to $0.0001*N$ after 2 warmup epochs.}

The last layer of the network has two softmax nodes, which provide the output probability that the input image belongs to a given galaxy class. To quantify the over-all accuracy of the neural network, we extract the output probabilities from this last layer for our two HP and FO test sets, and compare them against the ground truth labels provided through the Galaxy Zoo project. While these probabilities can be directly tested for cross-matched DES sets by comparing to the SDSS-Galaxy Zoo probabilities, for the rest of the unlabelled DES images this is not possible. For large-scale galaxy catalogs it would be unfeasible to inspect individual images to determine what class they belong to and check against the neural networks out-put probabilities to check for consistency. In practice, we can use the nodes of the second last layer of the neural network to determine what combination of them is activated for each galaxy type. In this approach, the activation vectors of this layer would form two distinct clusters, for each galaxy type in a 1024-D space. Checking whether similar combinations of neurons are activated, i.e., similar clusters are formed for the unlabelled DES data as the FO and HP test sets, will serve as a heuristic check that the learning is successfully transferred to the classification of unlabelled DES images for the purpose of constructing a catalog. For example, if there is a lack of distinct clusters, or more than two clusters are seen, then that would suggest unknown types that are forced into being classified as spiral or elliptical because the output layer has only two nodes.

In order to visualize these 1024-D clusters, we embed them into a 3-D parameter space using the sklearn library implementation of t-Distributed Stochastic Neighbor Embedding (t-SNE)~\cite{TSNE}. t-SNE is a nonlinear dimensionality reduction technique that is particularly apt for visualizing high-dimensional datasets by finding a faithful representation in a low dimensional embedding, typically 2-D or 3-D. It is important to note that t-SNE adjusts its notion of distance to regional density variations in the dataset, and hence bounding boxes of clusters in the low dimensional representation don't correspond to their relative sizes. Similarly, distances between clusters may not be meaningful since they are affected by a number of hyper-parameters such as perplexity and number of iterations.

As a final step, we introduce an application of unsupervised/semi-supervised learning in the form of recursive training, where we introduce into the training set newly labelled DES galaxies and retrain our model. It has been suggested in~\cite{Dom:2018D} that once trained with a particular dataset from one survey, neural networks can quickly adapt to new instrument
characteristics (e.g., PSF, seeing, depth), reducing by almost one order of magnitude
the necessary training sample from a different survey for morphological classification. However instead of manually labelling new DES images, we extract the out-put classification probabilities for them through our fine-tuned neural network. We use a sample of ~10,000 unlabelled bright DES galaxies and then by introducing a threshold on the neural networks classification confidence, we select the 1000 most confident predictions for spiral and elliptical respectively to further fine tune our network on.

%%%%%%%%%%%%%%%%%%%%%%%%%%%%%%%%%%%%%%%%%%%%%
%%%%%%%%%%%%%%%%%%%%%%%%%%%%%%%%%%%%%%%%%%%%%

%%%%%%%%%%%%%%%%%%%% results on cooley %%%%%%%%%%%%%%%%%%%%%%%%%
%%%%%%%%%%%%%%%%%%%%  %%%%%%%%%%%%%%%%%%%%%%%

\section{Results}
\label{Clustering}

To check the performance of our neural network model on HP and FO test sets, we use standard evaluation metrics: precision, recall, accuracy and F1 score. For binary classification, precision is the number of true positives divided by the total number of predicted positives, i.e., true positives plus false positives. Similarly, recall is the number of true positives divided by the total number of actual positives, i.e., true positives plus false negatives.  The F1 score is a single number statistical evaluation metric that measures the accuracy of binary classification by taking a weighted average of precision and recall. It varies between its worst performance value of 0 and best performance value of 1, and is given by

\beq
\textrm{F1\, score} =2\,\frac{\textrm{precision}\times \textrm{recall}}{\textrm{precision}+\textrm{recall}} \,.
\label{f1}
\eeq

\begin{figure*}[t!]
\centerline{
\includegraphics[width=0.45\linewidth, height=0.35\linewidth]{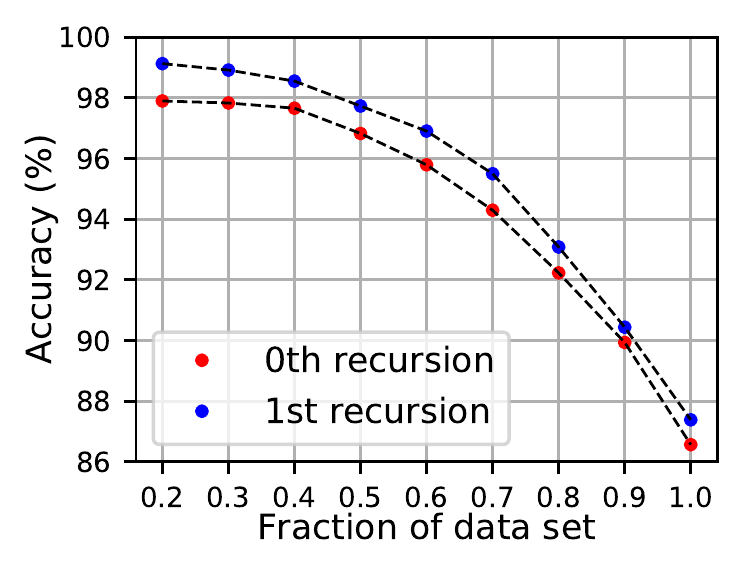}
\includegraphics[width=0.45\linewidth, height=0.35\linewidth]{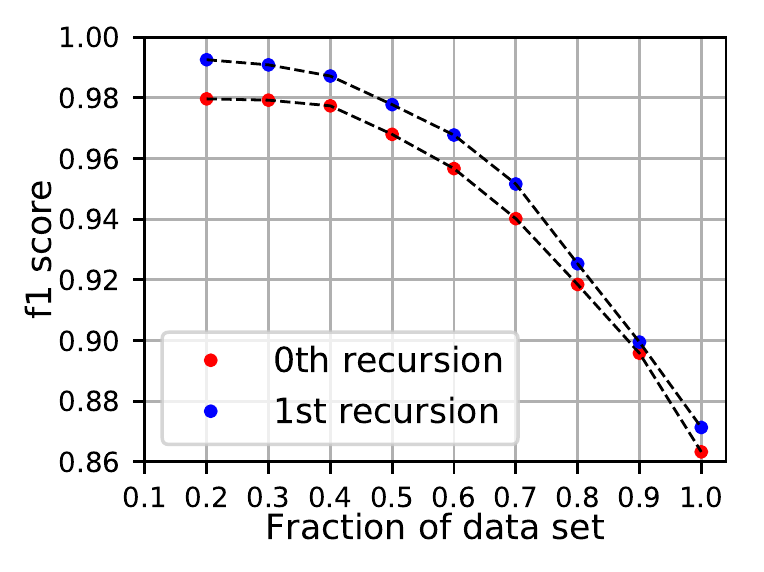}
}
\centerline{
\includegraphics[width=0.45\linewidth, height=0.35\linewidth]{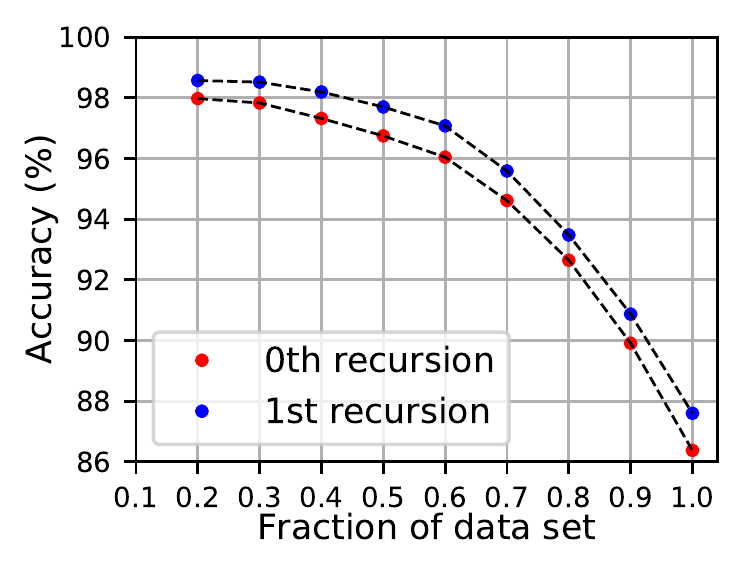}
\includegraphics[width=0.45\linewidth, height=0.35\linewidth]{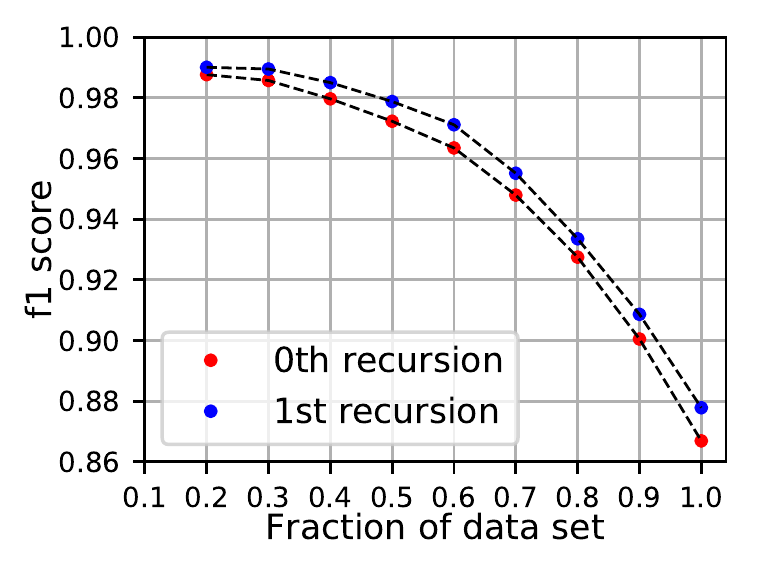}
}
\caption{Top panels: SDSS datasets. Bottom panels: DES datasets. Accuracy (left panels) and F1 score (right panels) vs N high confidence predictions as a fraction of total full overlap test datasets (0th recursion). We also show the improvement in classification accuracy and F1 score after 2000 newly labelled DES images are added to the SDSS training dataset (1st recursion). These results have been obtained by averaging over ten different models.}
  \label{fig:mask_size}
\end{figure*}

\noindent The performance of these metrics on the various test sets is summarized in Table~\ref{table:summary-results}. As can be seen in Table~\ref{table:summary-results}, deep transfer learning from everyday object classification in the \texttt{ImageNet} dataset to morphological classification of galaxies in SDSS and DES leads to state of the art accuracies and F1 scores. Our fine tuned \texttt{Xception} model attains accuracies \(\gtrsim 99\%\) for the HP SDSS and HP DES test sets. Unlike HP test sets, the FO test sets do not entirely consist of galaxies with robust ground truth classifications. Hence, instead of applying a threshold on the ground truth probabilities, we apply a confidence threshold on the predictions of the neural network. In Table~\ref{table:summary-results} we pick the top half most confident predictions, for which the neural network attains \(\gtrsim96\%\) accuracies and F1 scores. Additionally, the accuracies and F1 scores obtained by applying various different thresholds are also summarized in Fig~\ref{fig:mask_size}, while the Receiver Operating Characteristic (ROC) curves are shown in Fig~\ref{fig:roc_recursive}~\ref{ap5}.

%%%%%%%%%%%%%%%%%%%% DATA SUMMARY %%%%%%%%%%%%%%%%%%%%%%%%%
\begin{table}[h!]
		\footnotesize
		\begin{center}
                        \setlength{\tabcolsep}{3pt} % default is apparently 6pt
			\begin{tabular}{|c c c c c c|}
				\hline 
				Dataset & Precision & Recall & FPR & Accuracy & F1 score \\ 
				\hline
				Training set  & & & & 99.81\% & 0.9998 \\
				 \hline
				HP SDSS Test Set & 0.996 & 1 & 0.004 &  99.81\% & 0.9980 \\
				HP DES\,\,\,   Test Set & 0.998 & 0.995 & 0.002 &  99.62\% & 0.9961 \\
				\hline
				FO SDSS Test Set & 0.945 & 0.991 & 0.055 &  96.76\% & 0.9675 \\
				FO DES\,\,\,   Test Set & 0.965 & 0.946 & 0.025 &  96.32\% & 0.9685 \\
				\hline
			\end{tabular}
		\end{center}
	\caption{Classification accuracy for each test dataset.}
	\label{table:summary-results}
	\end{table}
	\normalsize
%%%%%%%%%%%%%%%%%%%% Prob Dist of each Data set %%%%%%%%%%%%%%%%%%%%%%%

\noindent Having quantified the accuracy of our neural network model on DES test sets that overlap the SDSS footprint, we now feed our model with bright, unlabelled DES galaxies that do not overlap the SDSS footprint, and predict their classes, thereby labelling them. A random sample of high confidence predictions for these is shown in Figure~\ref{fig:des_pred} in~\ref{ap3}.

\begin{figure*}[h!]
\centerline{
\raisebox{3mm}{\includegraphics[width=0.3\linewidth, height=0.31\linewidth]{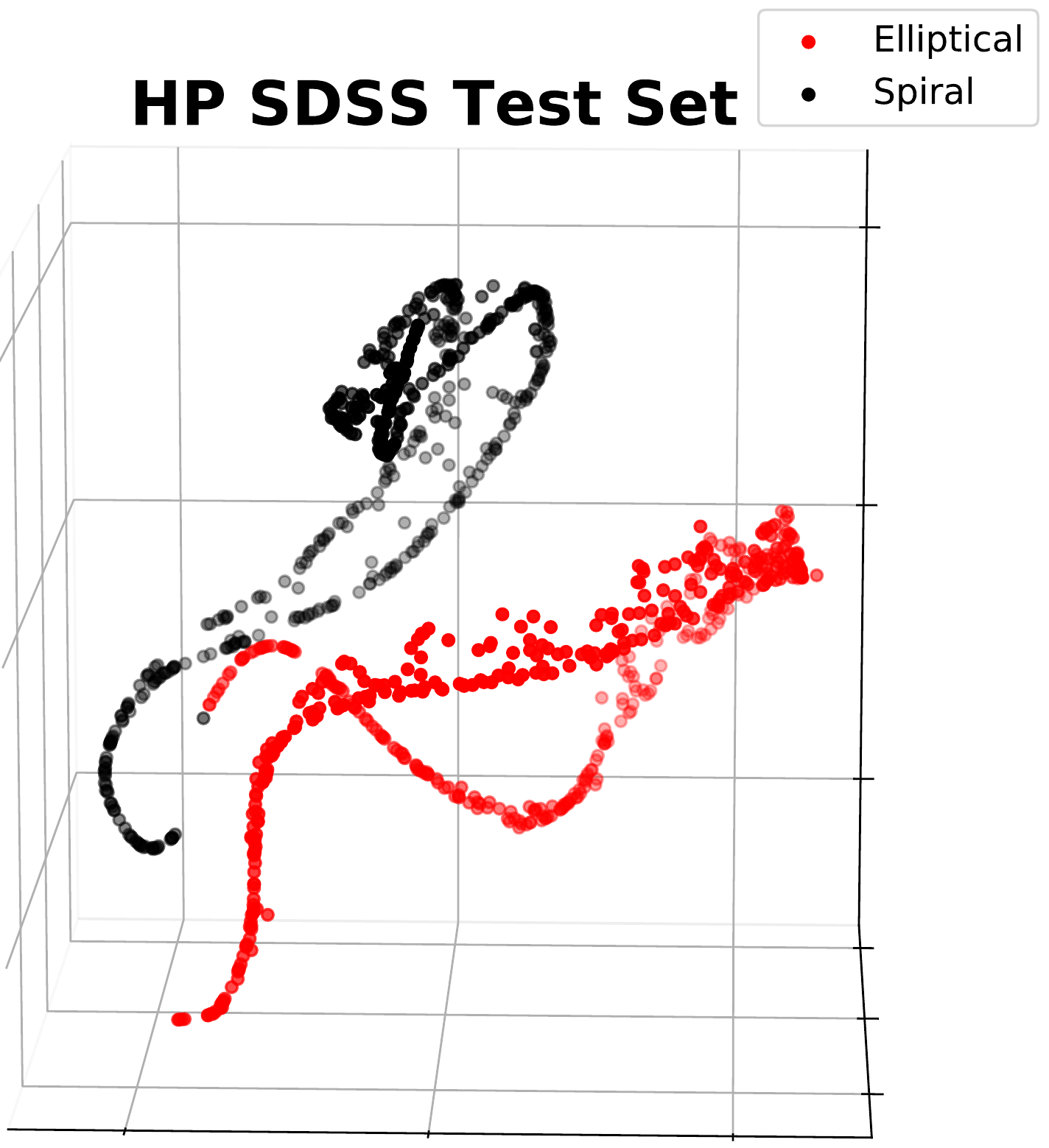}}
\raisebox{1.5mm}{\includegraphics[width=0.3\linewidth, height=0.31\linewidth]{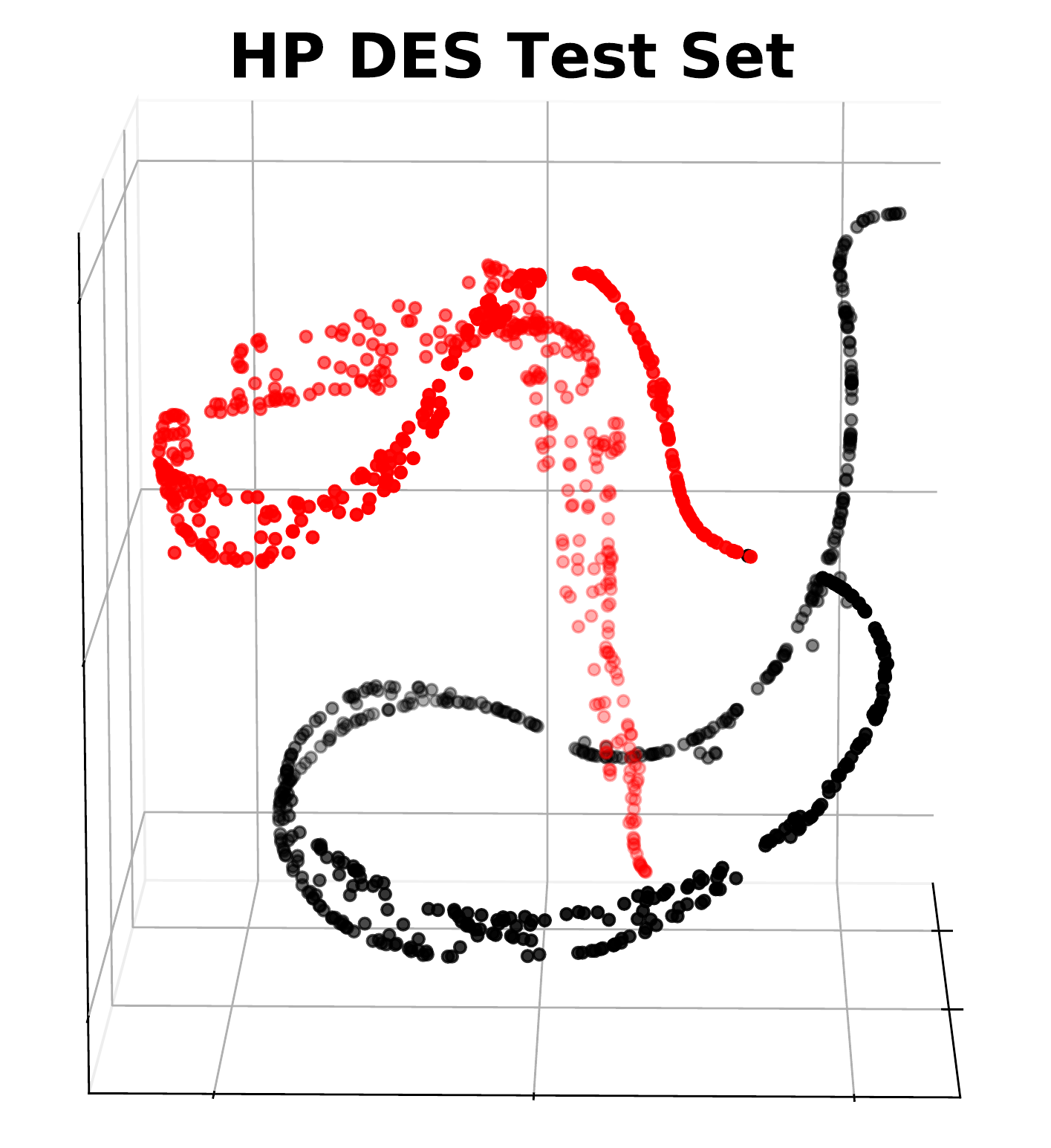}}
\includegraphics[width=0.3\linewidth, height=0.31\linewidth]{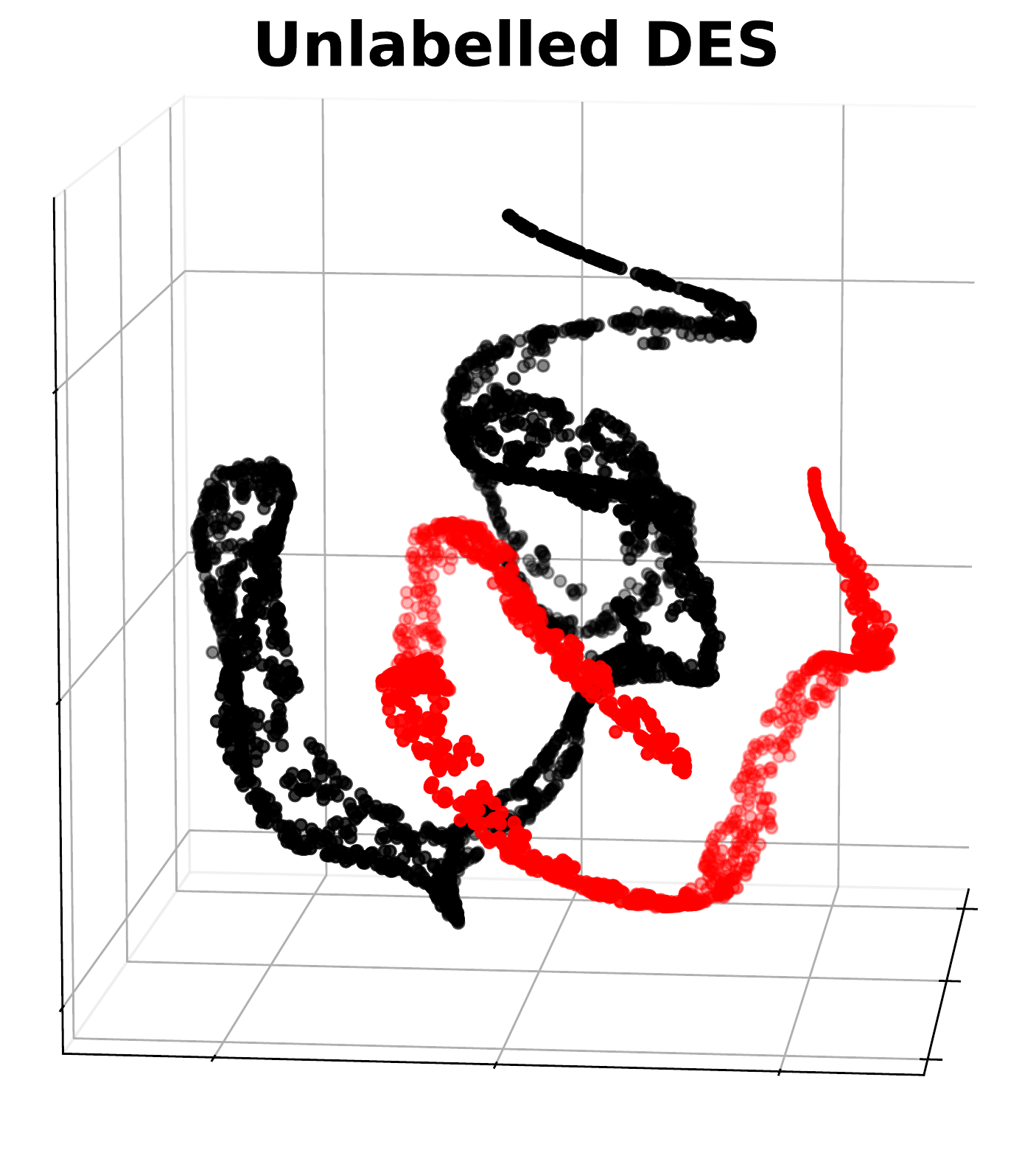}
} 
\caption{t-SNE visualization of the clustering of HP SDSS and DES test sets, and unlabelled DES test.}
\label{fig:tSNE}
\end{figure*}

%%%%%%%%%%%%%%%%%%%% Accuracy vs Confidence %%%%%%%%%%%%%%%%%%%%%%%%%

\noindent Lastly, using our neural network model as a feature extractor, we obtain the activation maps of the second last layer and embed them in 3-D using t-SNE. The left and middle panels of Figure~\ref{fig:tSNE} show the output of t-SNE when applied to the HP SDSS and HP DES test sets. We labelled the points using the ground-truth label of each galaxy, and found that the points neatly cluster into two groups with accuracies \(\gtrsim 99\%\). For unlabelled DES set, shown in the right panel of Figure~\ref{fig:tSNE}, we find again that two distinct clusters are formed. Based on the accuracy of the FO DES test set, we heuristically know that these clusters have accuracies \(\gtrsim96\%\) for the top-half most confident predictions. These results indicate that the neural network model has extracted the necessary information from the training dataset to enable t-SNE to clearly identify two distinct classes of galaxies based on their morphology. A scientific visualization of this clustering algorithm for the FO DES test set is presented in~\cite{viztsne}. The astute reader may realize 
that using t-SNE as a visualization tool requires training to prevent common misreadings of the 
visualizations. Furthermore, t-SNE does not always produce similar outputs on successive runs, 
and it requires the user to determine a few hyperparamters related to the optimization process. 
Much work has been presented in the literature to ensure that new users make a proper use of this 
tool~\cite{wattenberg2016how}, and to automate hyperparamter selection~\cite{cao_perplex_2017}.

\noindent {\bf Recursive training} Having labeled about 10,000 DES galaxies with our neural network classifier, we pick the top 1000 spiral and top 1000 elliptical galaxies. We then add them to our original SDSS training dataset, and use 
deep transfer learning again to re-train the neural network model. The top- and bottom-left panels in Figure~\ref{fig:mask_size} show the initial (0th recursion) accuracy of our classifier, and the accuracy attained 
once the newly labelled DES images are added to the SDSS training dataset (1st recursion). We notice that 
the accuracy for classification for FO SDSS and DES test sets improves up to \(1.5\%\). In particular, we notice that the classification accuracy for the FO DES test set is now boosted 
up to \(98.5\%\) when 50\% of the dataset is considered. These results are rather significant from a machine learning perspective~\cite{mnist_bm}, since these accuracies are already high and this newly labelled DES dataset represents \(\sim 5\%\) of the the original SDSS training dataset. We have also computed ROC curves (see Figure~\ref{fig:roc_recursive}) to provide an additional metric to quantify the improvement in classification accuracy due to recursive training.

Intuitively, recursive training provides the means to continually enhance the classification accuracy of a neural network as new data becomes available. We have found that, averaging 
over ten models, the mean classification accuracies do improve when we retrain the model.

\noindent This novel approach provides us with the means to enhance SDSS galaxy classification, as shown 
in the top left panel of Figure~\ref{fig:mask_size}. More importantly, it provides a way forward to gradually replace SDSS galaxy images in the training dataset that we need to construct DES galaxy catalogs at scale. A DES-only image 
training dataset will better capture the nature of images observed by DES, and would also enable us to better use data augmentations to model the effects of noise, making our neural network model more 
resilient to accurately classify galaxies at higher redshift, or that are contaminated by various sources of noise. 

%%%%%%%%%%%%%%%%%%%%%%%%%%%%%%%%%%%%%%%%%%%%%%%%%%%%%%
%%%%%%%%%%%%%%%%%%%%%%%%%%%%%%%%%%%%%%%%%%%%%%%%%%%%%%
\section{Conclusion}
\label{end}
We have presented the first application of deep transfer learning combined with distributed training for the classification of 
DES galaxies that overlap the footprint of the SDSS survey, achieving state-of-the-art 
accuracies \(\gtrsim 99.6\%\). We described how to use our neural network classifier to label over 10,000 unlabelled DES galaxies that had not been observed in previous surveys. By truncating our neural network model, we used it as a feature extractor, and once combined with t-SNE, we presented visualizations which show that, through transfer learning, the neural network abstracted morphological information to clearly identify two distinct classes of galaxies in the unlabeled DES dataset. To get insights into the inner workings of our clustering algorithm, we 
have presented scientific visualizations of the clustering of the FO DES test set, which are 
available at~\cite{viztsne} and~\cite{viztsne2} .

We have also used t-SNE to inspect seemingly incorrect labels 
provided by our neural network model, and have found that these errors actually correspond to inaccurate human classifications in our SDSS testing dataset. We present an example of this nature in the visualization available at ~\cite{viztsne}, and in Figure~\ref{fig:misclassification}.

Furthermore, adding the most confident predictions from our newly labeled DES galaxies, we have done recursive training, boosting the classification accuracy for the FO SDSS and DES test datasets. Averaging over ten models, we find improved accuracies as high as 99.5\% for SDSS and 99\% for DES.

This analysis provides a path forward to construct galaxy catalogs in DES using actual DES galaxies as training datasets. The combination of deep transfer learning with distributed training, and recursive training presents an alternative to do this analysis at scale in the LSST era.

%%%%%%%%%%%%%%%%%%%%%%%%%%%%%%%%%%%%%%%%%%%%%
%%%%%%%%%%%%%%%%%%%%%%%%%%%%%%%%%%%%%%%%%%%%%
\section{Acknowledgements}
\label{ack}

This research is part of the Blue Waters sustained-petascale computing project, 
which is supported by the National Science Foundation (awards OCI-0725070 and ACI-1238993) 
and the State of Illinois. Blue Waters is a joint effort of the University of Illinois at 
Urbana-Champaign and its National Center for Supercomputing Applications. 
We acknowledge support from the NCSA. We thank the 
\href{http://gravity.ncsa.illinois.edu}{NCSA Gravity Group} for useful feedback, 
and Vlad Kindratenko for granting us access to state-of-the-art GPUs and HPC 
resources at the Innovative Systems Lab at NCSA. We are grateful to NVIDIA 
for donating several Tesla P100 and V100 GPUs that we used for our analysis. We thank the anonymous referee for carefully reading this manuscript, and providing constructive feedback to improve the presentation of our results.

This work used the Extreme Science and Engineering Discovery Environment (XSEDE), which is supported by National Science Foundation grant number ACI-1548562. Specifically, it used the Bridges system, which is supported by NSF award number ACI-1445606, at the Pittsburgh Supercomputing Center (PSC). We gratefully acknowledge grant TG-PHY160053. This research used resources of the Argonne Leadership Computing Facility, which is a DOE Office of Science User Facility supported under Contract DE-AC02-06CH11357.

This project used public archival data from the Dark Energy Survey (DES). 
Funding for the DES Projects has been provided by the U.S. Department of 
Energy, the U.S. National Science Foundation, the Ministry of Science and 
Education of Spain, the Science and Technology Facilities Council of the United 
Kingdom, the Higher Education Funding Council for England, the National Center 
for Supercomputing Applications at the University of Illinois at Urbana-Champaign, 
the Kavli Institute of Cosmological Physics at the University of Chicago, the Center 
for Cosmology and Astro-Particle Physics at the Ohio State University, the Mitchell 
Institute for Fundamental Physics and Astronomy at Texas A\&M University, 
Financiadora de Estudos e Projetos, Funda\c{c}\~{a}o Carlos Chagas Filho de 
Amparo {\`a} Pesquisa do Estado do Rio de Janeiro, Conselho Nacional de 
Desenvolvimento Cient\'ifico e Tecnol\'ogico and the Minist\'erio da Ci\^{e}ncia, 
Tecnologia e Inova\c{c}\~{a}o, the Deutsche Forschungsgemeinschaft and 
the Collaborating Institutions in the Dark Energy Survey.

The Collaborating Institutions are Argonne National Laboratory, 
the University of California at Santa Cruz, the University of Cambridge, 
Centro de Investigaciones Energ\'eticas, Medioambientales y Tecnol\'ogicas--Madrid, 
the University of Chicago, University College London, the DES-Brazil Consortium, 
the University of Edinburgh, the Eidgen{\"o}ssische Technische Hochschule (ETH) Z{\"u}rich, 
Fermi National Accelerator Laboratory, the University of Illinois at Urbana-Champaign, 
the Institut de Ci{\`e}ncies de l'Espai (IEEC/CSIC), the Institut de F\'isica d'Altes Energies, 
Lawrence Berkeley National Laboratory, the Ludwig-Maximilians Universit{\"a}t 
M{\"u}nchen and the associated Excellence Cluster Universe, the University of Michigan, 
the National Optical Astronomy Observatory, the University of Nottingham, The 
Ohio State University, the OzDES Membership Consortium, the University of 
Pennsylvania, the University of Portsmouth, SLAC National Accelerator Laboratory, 
Stanford University, the University of Sussex, and Texas A\&M University.
Based in part on observations at Cerro Tololo Inter-American Observatory, 
National Optical Astronomy Observatory, which is operated by the Association 
of Universities for Research in Astronomy (AURA) under a cooperative 
agreement with the National Science Foundation.

%%%%%%%%%%%%%%%%%%%%%%%%%%%%%%%%%%%%%%%%%%%%%
%%%%%%%%%%%%%%%%%%%%%%%%%%%%%%%%%%%%%%%%%%%%%
\bibliography{references}
\bibliographystyle{apsrev4-1}
%%%%%%%%%%%%%%%%%%%%%%%%%%%%%%%%%%%%%%%%%%%%%
%%%%%%%%%%%%%%%%%%%%%%%%%%%%%%%%%%%%%%%%%%%%%
%\clearpage

\appendix

\clearpage

\begin{widetext}

\section{Neural network architecture}
\label{ap1}

    The top panel in Figure~\ref{fig:model} presents the architecture of the pre-trained \texttt{Xception} model~\cite{chollet:2016} used in these studies. The bottom panel of Figure~\ref{fig:model} shows the fully connected layers, and classifier added at the second-to-last layer of the pre-trained \texttt{Xception} model. This is labeled as \texttt{2048-dimensional vectors} in the \texttt{Exit flow} diagram.  Following this procedure, we truncate our neural network classifier and turn it into a feature extractor, which we can use in combination with t-SNE to do unsupervised clustering. Note that we use t-SNE just as a visual aid. The labelling of unlabeled DES datasets is done using our neural network classifier.

\begin{figure*}[htb!]
%\centerline{
\includegraphics[width=\linewidth]{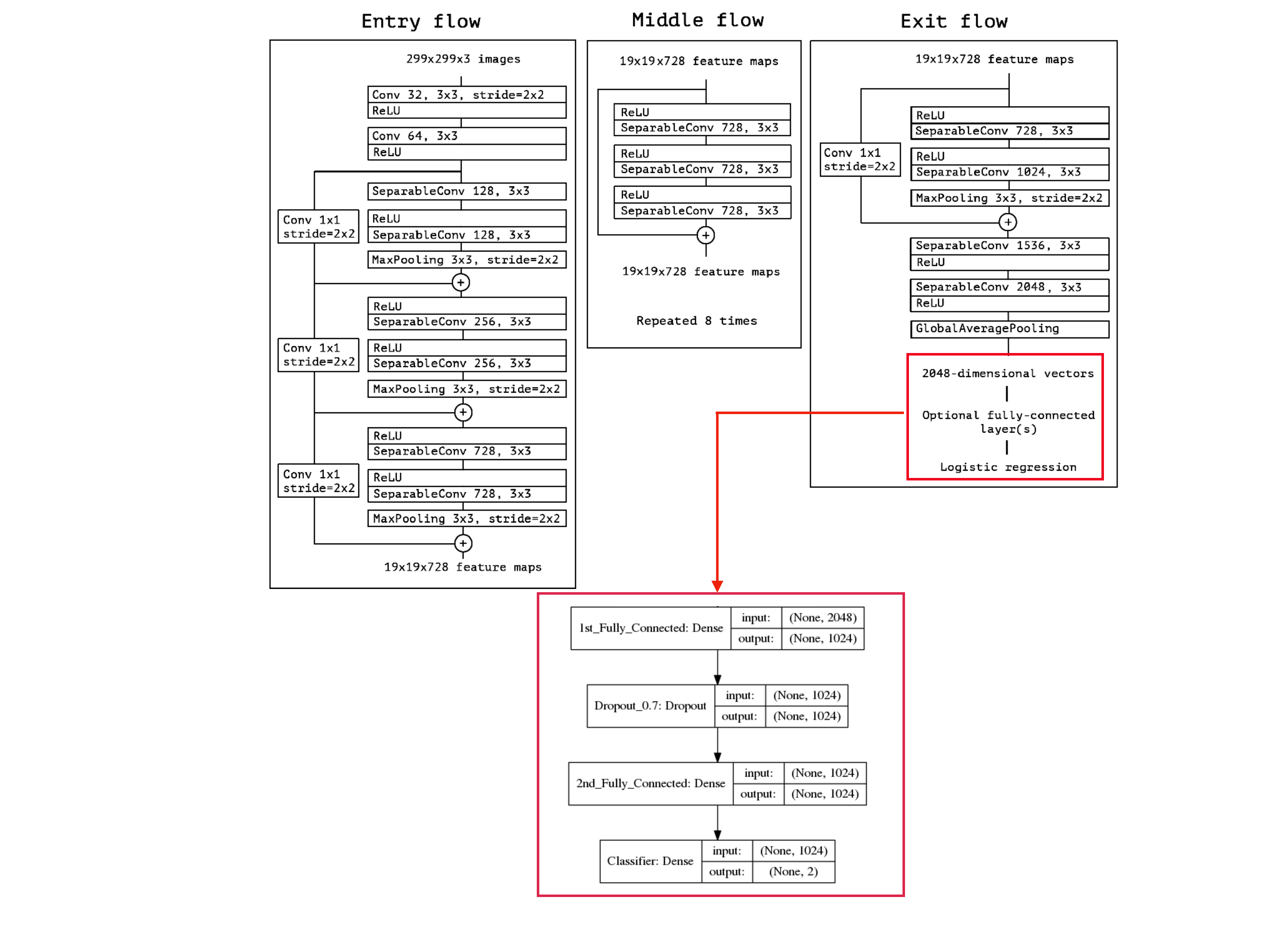}
%}
\caption{Top panel: \texttt{Xception} model~\cite{chollet:2016}. Bottom panel: fully connected layers, and classifier added at the bottleneck of the pre-trained \texttt{Xception} model.}
\label{fig:model}
\end{figure*}

%%%%%%%%%%%%%%%%%%%%%%%%%%%%%%%%%%%%%%%%%%%%%
%%%%%%%%%%%%%%%%%%%%%%%%%%%%%%%%%%%%%%%%%%%%%

\section{Data Augmentation}
\label{ap2}

\begin{figure*}[htb!]
\centerline{
\includegraphics[width=0.4\linewidth]{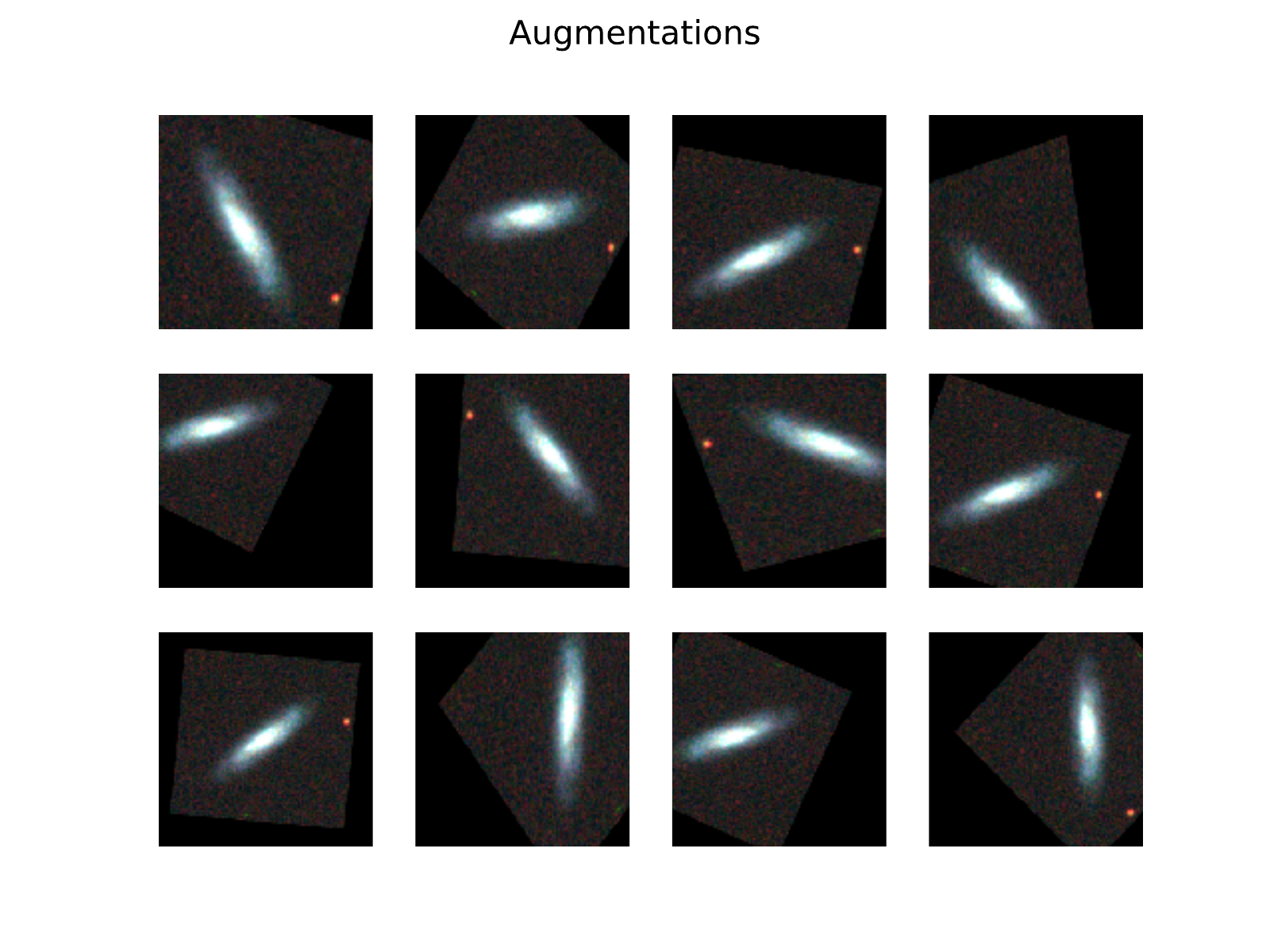}
}
\caption{Data augmentations include random rotations of up to 45 degrees, random flips, height and width shifts and zooms of up to a factor of 1.3}
\label{fig:data_augmentation}
\end{figure*}

To expose the neural network to a variety of potential scenarios for classification, we augment original galaxy 
images with random vertical and horizontal flips, random rotations, height and width shifts and zooms, as shown in
Figure~\ref{fig:data_augmentation}. The range for random rotations is set up to a maximum of 45$^{\circ}$, i.e. for each iteration of training the neural net sees the input image rotated randomly between -45$^{\circ}$ and 45$^{\circ}$.  Similarly the maximum range for random height and width shifts, as well as zooming factor is set to 0.3. Note that for each iteration of training all these image transformations are applied but with random values within the defined ranges.

This approach not only synthetically increases the training dataset, but also makes the neural network invariant to rotations, shifts, flips and combinations of these, and also introduces 
scale invariance.

%%%%%%%%%%%%%%%%%%%%%%%%%%%%%%%%%%%%%%%%%%%%%
%%%%%%%%%%%%%%%%%%%%%%%%%%%%%%%%%%%%%%%%%%%%%

\section{Classification predictions for unlabelled DES galaxies}
\label{ap3}

Figure~\ref{fig:des_pred} presents high-confidence neural network predictions for unlabelled DES galaxies. 
The robustness of these predictions were tested with our unsupervised clustering algorithm, finding 
that these classifications, based on the morphological features extracted from the DES images in three filters, 
are meaningful, as shown in the t-SNE projections in Figure~\ref{fig:tSNE}. 

\begin{figure*}[htb!]
\centerline{
\includegraphics[width=0.47\linewidth]{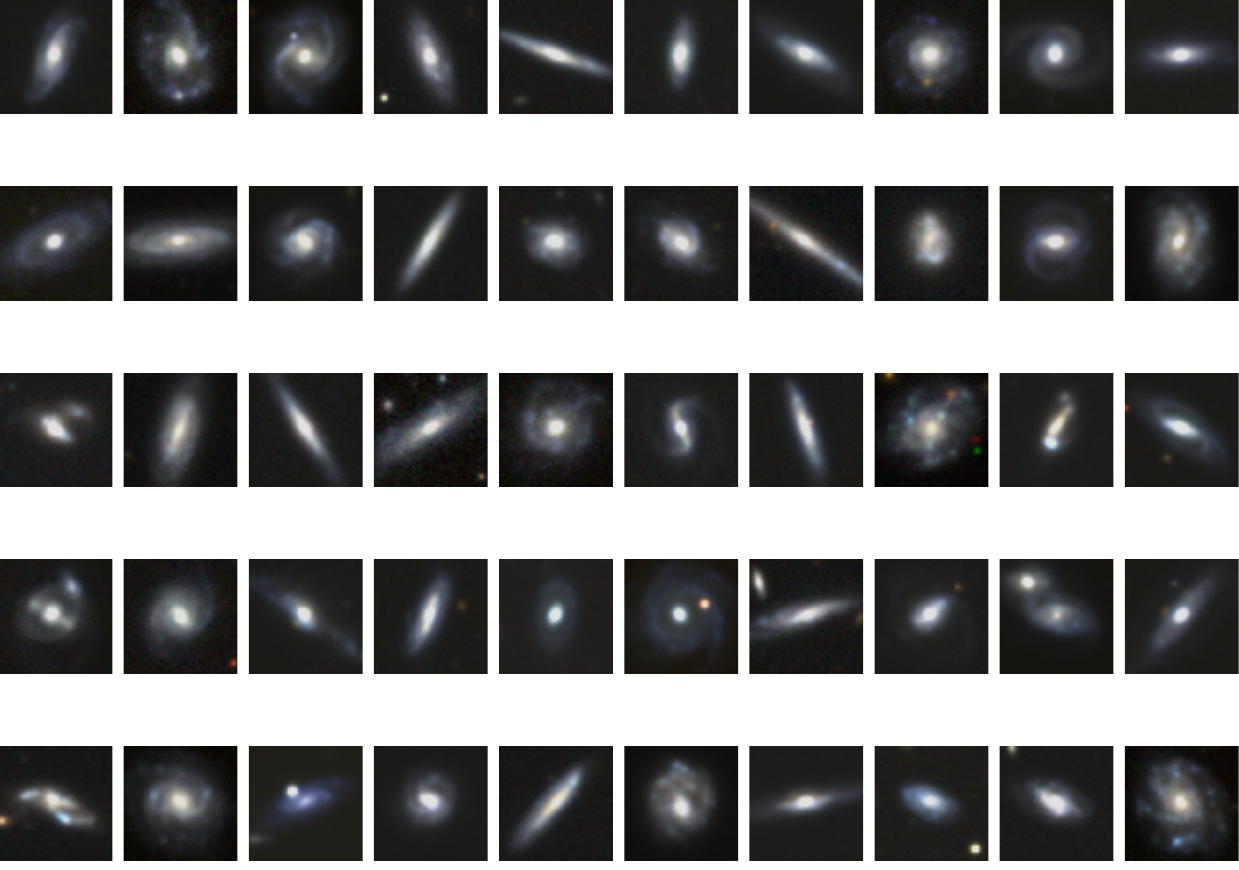}\hspace{1em}
\raisebox{0.4mm}{%
\includegraphics[width=0.472\linewidth]{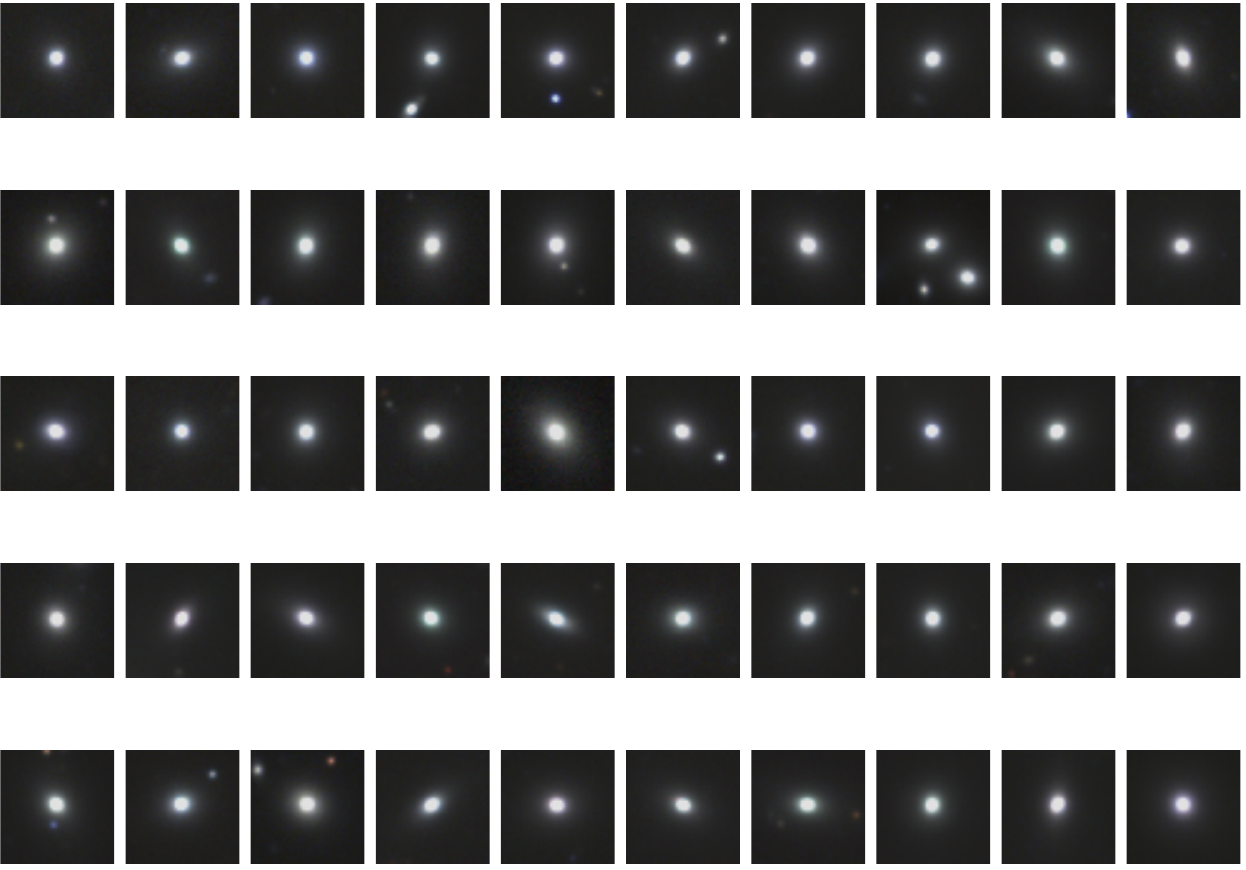}
}
}
\caption{Sample of high confidence predictions for spiral (left panel) and elliptical galaxies (right panel) on an unlabelled DES set.}
\label{fig:des_pred}
\end{figure*}

\section{Scaling Results}
\label{ap4}
In Table\ref{table:summary-scaling-results} the training is comprised of three stages: (1) freeze the base model and train added dense
layers; (2) freeze layers 0-39, and train layers 40+; (3) freeze layers 0-1 and train all layers 2+. The number of epochs for each stage is given in the third column. The total time only includes the time for the training (all three stages), but not the
time for initialization (launching jobs, loading Python modules, data preparation, etc), which we found was minimal compared to the training time.

%%%%%%%%%%%%%%%%%%%% results on cooley %%%%%%%%%%%%%%%%%%%%%%%%%
\begin{table}[h!]
		\footnotesize 
		\begin{center}
                        \setlength{\tabcolsep}{7pt} % default is apparently 6pt
			\begin{tabular}{|c|c| c| c| c| c| c |}
				\hline 
				 GPUs &  Time per epoch (s) & \# epochs & Total time & Accuracy & Val Accuracy \\ 
				\hline
				1  & \begin{tabular}{c c c}410 \\ 922 \\ 1626 \end{tabular} & \begin{tabular}{c c c}1 \\ 11 \\ 4 \end{tabular} & 4h 44 m& 0.9992 &0.9979 \\
				 \hline
				2 & \begin{tabular}{c c c}231 \\ 481 \\ 830 \end{tabular} & \begin{tabular}{c c c}1 \\ 6 \\ 4 \end{tabular} & 1h 47m & 0.9993 & 0.9990 \\
				\hline
				4 &\begin{tabular}{c c c}119 \\ 246 \\ 427 \end{tabular} & \begin{tabular}{c c c}1 \\ 5 \\ 7 \end{tabular} & 1h 12m & 0.9995 &0.9990 \\
				\hline
				8 &\begin{tabular}{c c c}64 \\ 124 \\ 214 \end{tabular} &\begin{tabular}{c c c}1 \\ 6 \\ 8 \end{tabular} &42m &0.9991 &0.9979\\
				\hline
				16 &\begin{tabular}{c c c}35 \\ 63 \\ 109 \end{tabular} &\begin{tabular}{c c c}1 \\ 4 \\ 17 \end{tabular} & 36m & 0.9993 &0.9980 \\
			    \hline
			    32 &\begin{tabular}{c c c}20 \\ 31 \\ 53 \end{tabular} &\begin{tabular}{c c c}1 \\ 6 \\ 12 \end{tabular} &14m &0.9993& 0.9990 \\
			    \hline
			    64 & \begin{tabular}{c c c}13 \\ 15 \\ 27 \end{tabular} &\begin{tabular}{c c c}1 \\ 5 \\ 15 \end{tabular} &8m&0.9993& 0.9990 \\
				\hline
			\end{tabular}
		\end{center}
	\caption{Training results and timings using different numbers of K80 GPUs for the Xception model. Validation Accuracy results are given in the final column.  The training includes three stages: (1) train the dense layers (base layers are frozen); (2) train layer 40+ (layers 0-39 are frozen); (3) train Layer 2+ (layer 0-1 are frozen). The number of epochs shown are for each training stage. The batch size is set to be 16. The benchmarks was done on Cooley supercomputer at Argonne Leadership Computer Facility (https://www.alcf.anl.gov/user-guides/cooley).}
	\label{table:summary-scaling-results}
	\end{table}
	\normalsize
%%%%%%%%%%%%%%%%%%%%  %%%%%%%%%%%%%%%%%%%%%%%

%%%%%%%%%%%%%%%%%%%%%%%%%%%%%%%%%%%%%%%%%%%%%
%%%%%%%%%%%%%%%%%%%%%%%%%%%%%%%%%%%%%%%%%%%%%

\section{Recursive Training}
\label{ap5}

In addition to providing results for recursive training using ten different models (see Figure~\ref{fig:mask_size}), herein we also provide ROC results for a typical model out of our ten samples. These ROC results indicate that recursive training indeed leads to an increase in classification accuracy both for SDSS and DES.

%%%%%%%%%%%%%%%%%%%% ROC-AUC Curves %%%%%%%%%%%%%%%%%%%%%%%%%
\begin{figure*}[t!]
\centerline{
\includegraphics[width=0.6\linewidth]{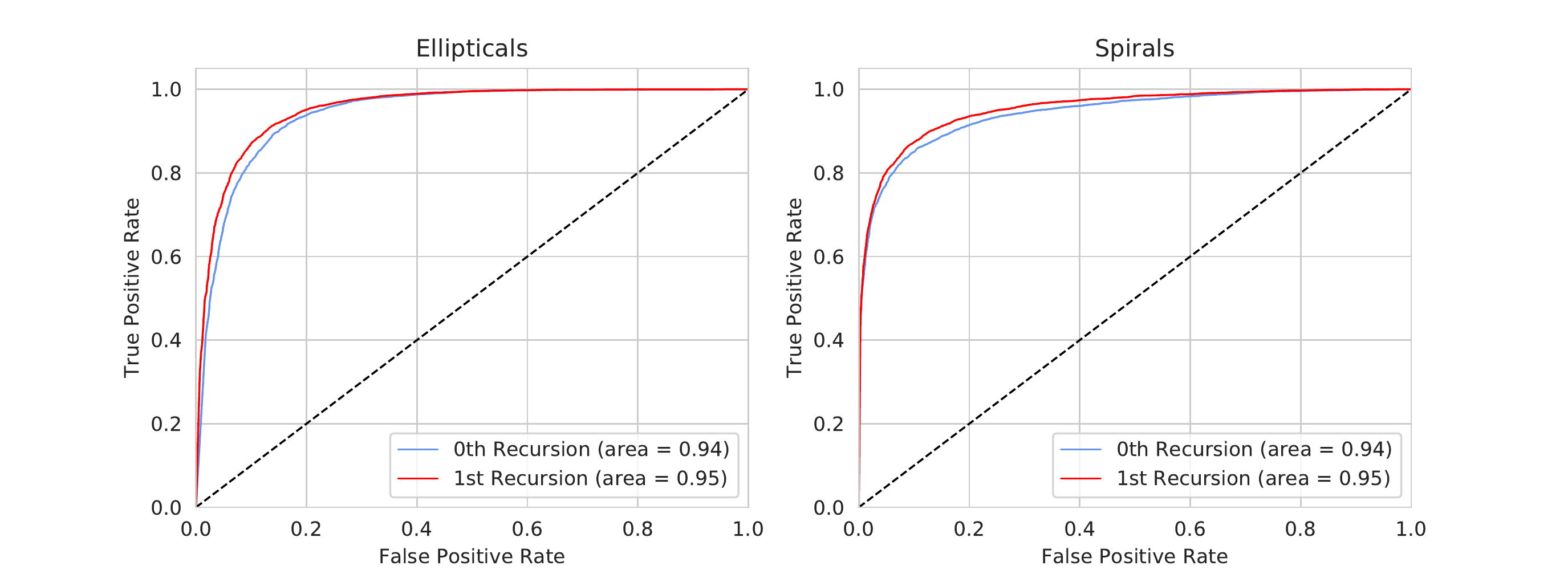}
}
\centerline{
\includegraphics[width=0.6\linewidth]{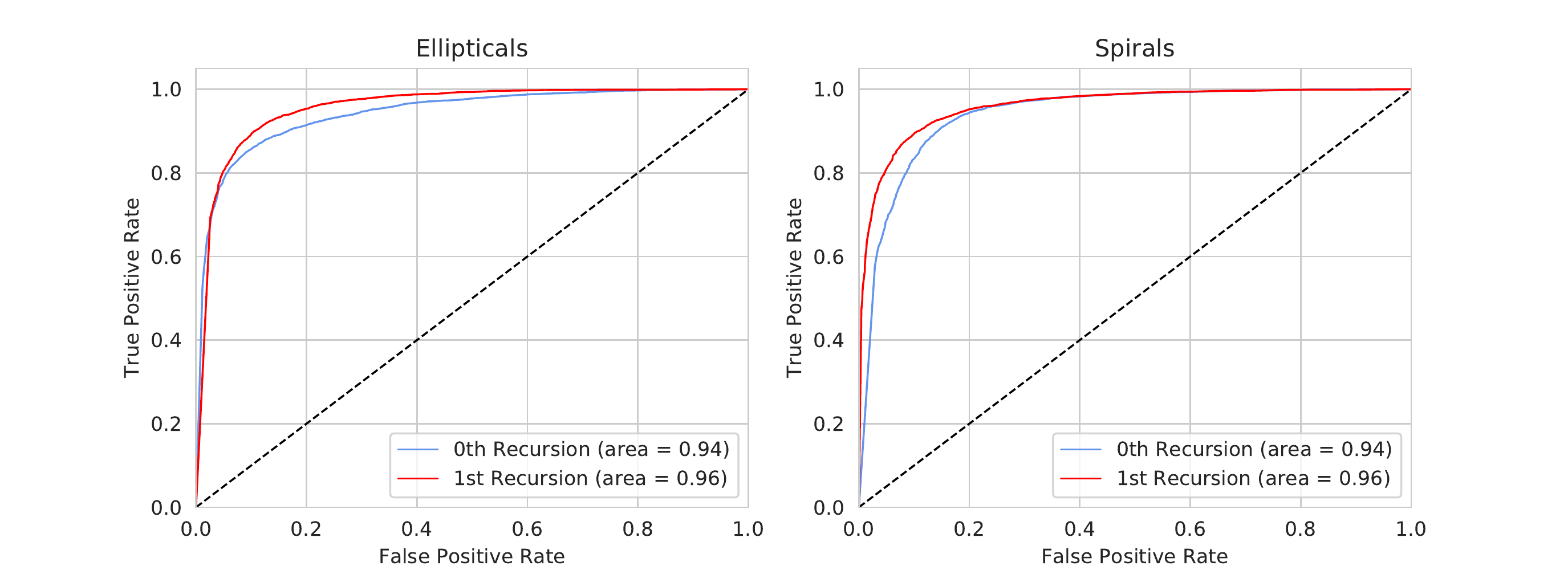}
}
\caption{Top panels: FO SDSS test sets. Bottom panels: FO DES test sets.}
  \label{fig:roc_recursive}
\end{figure*}

%%%%%%%%%%%%%%%%%%%%%%%%%%%%%%%%%%%%%%%%%%%%%
%%%%%%%%%%%%%%%%%%%%%%%%%%%%%%%%%%%%%%%%%%%%%

\section{Misclassified Examples}
\label{ap6}

We present a gallery of misclassified examples from our high probability (HP) test sets. As shown in Figure~\ref{fig:failed_samples}, we have only four instances of this nature, all from HP DES test set, and one of these corresponds to a noise artifact in the telescope.

%%%%%%%%%%%%%%%%%%%% HP misclassification %%%%%%%%%%%%%%%%%%%%%%%%%
\begin{figure*}[t!]
\centerline{
\includegraphics[width=0.35\linewidth]{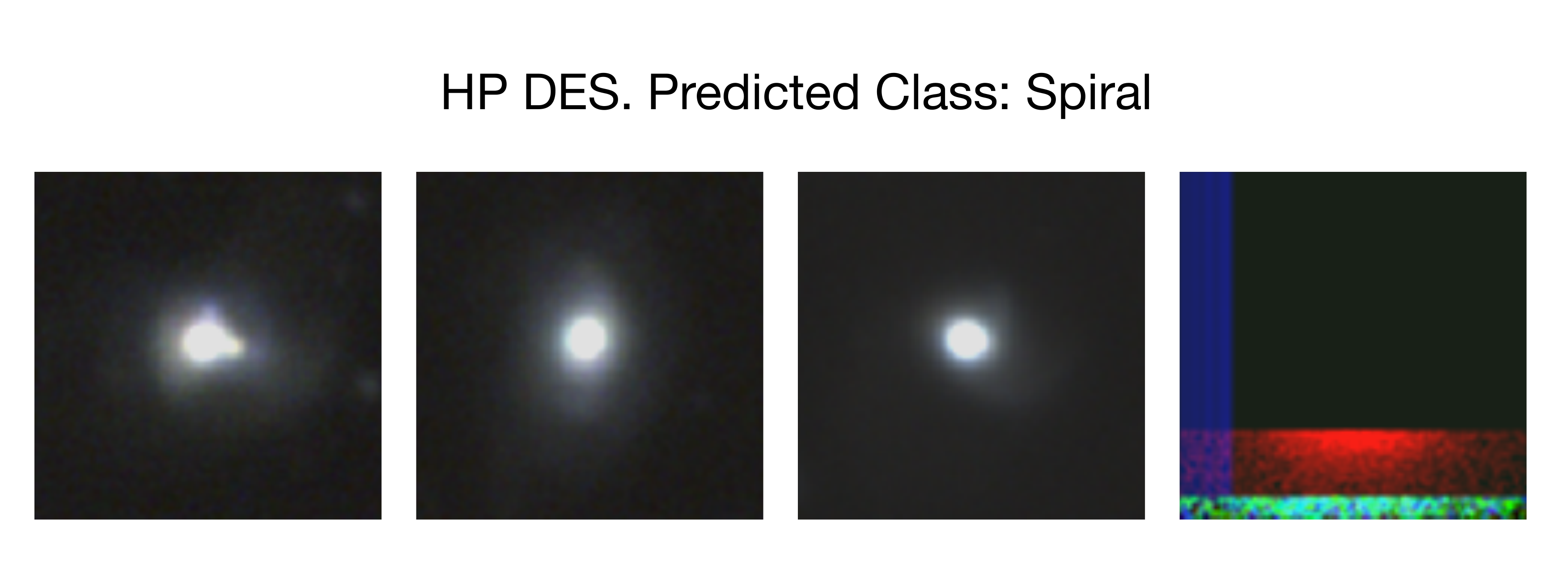}
}
\caption{There are only four misclassified examples from HP test sets. Of these four images, one is a noise artifact in the telescope.}
  \label{fig:failed_samples}
\end{figure*}

In Figure~\ref{fig:misclassification} we present a sample of inaccurate predictions on the full overlap (FO) test sets.

%%%%%%%%%%%%%%%%%%%% FO misclassification %%%%%%%%%%%%%%%%%%%%%%%%%
\begin{figure*}[t!]
\centerline{
\includegraphics[width=0.35\linewidth, height=0.25\linewidth]{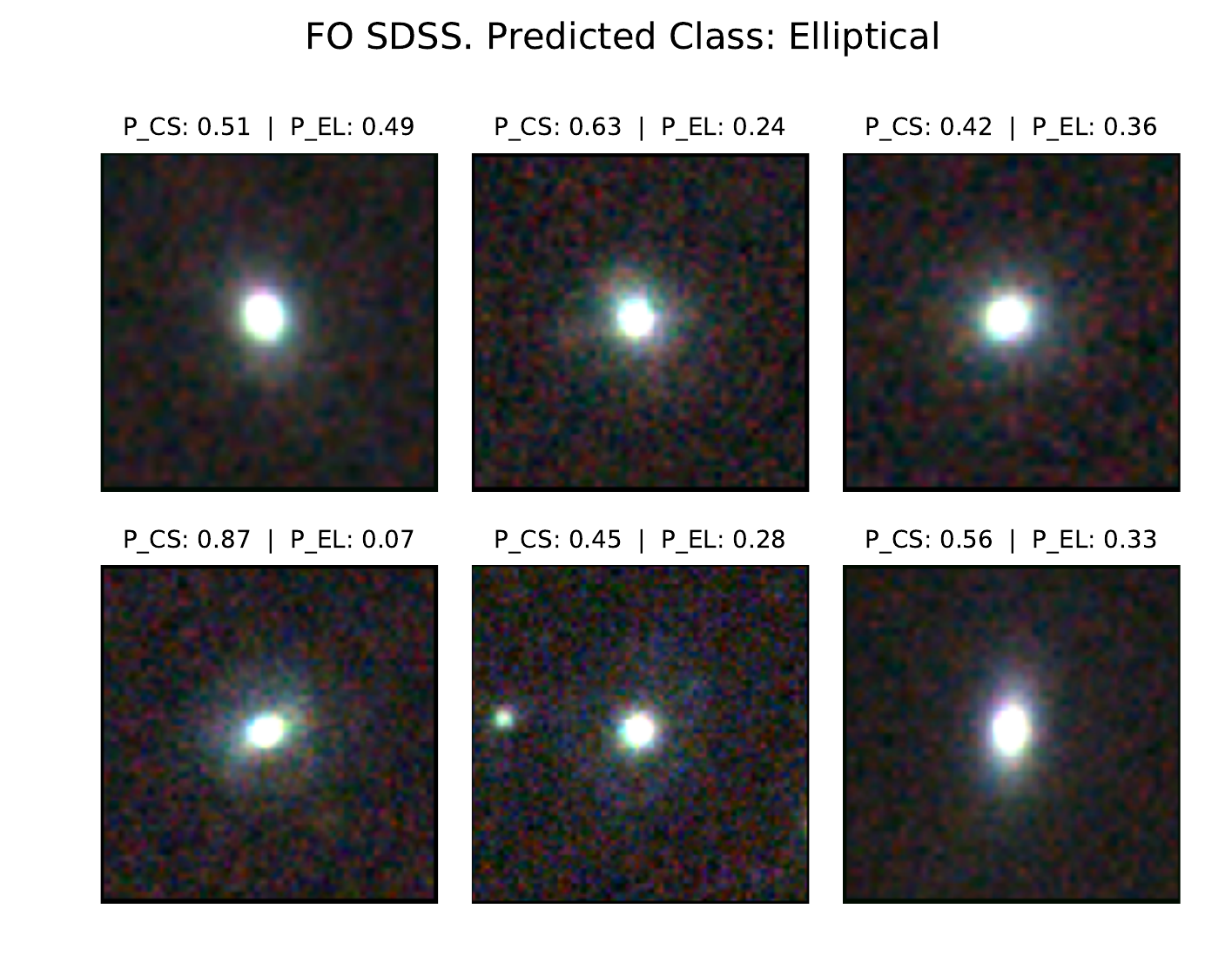}\hspace{-6mm}
\includegraphics[width=0.35\linewidth, height=0.25\linewidth]{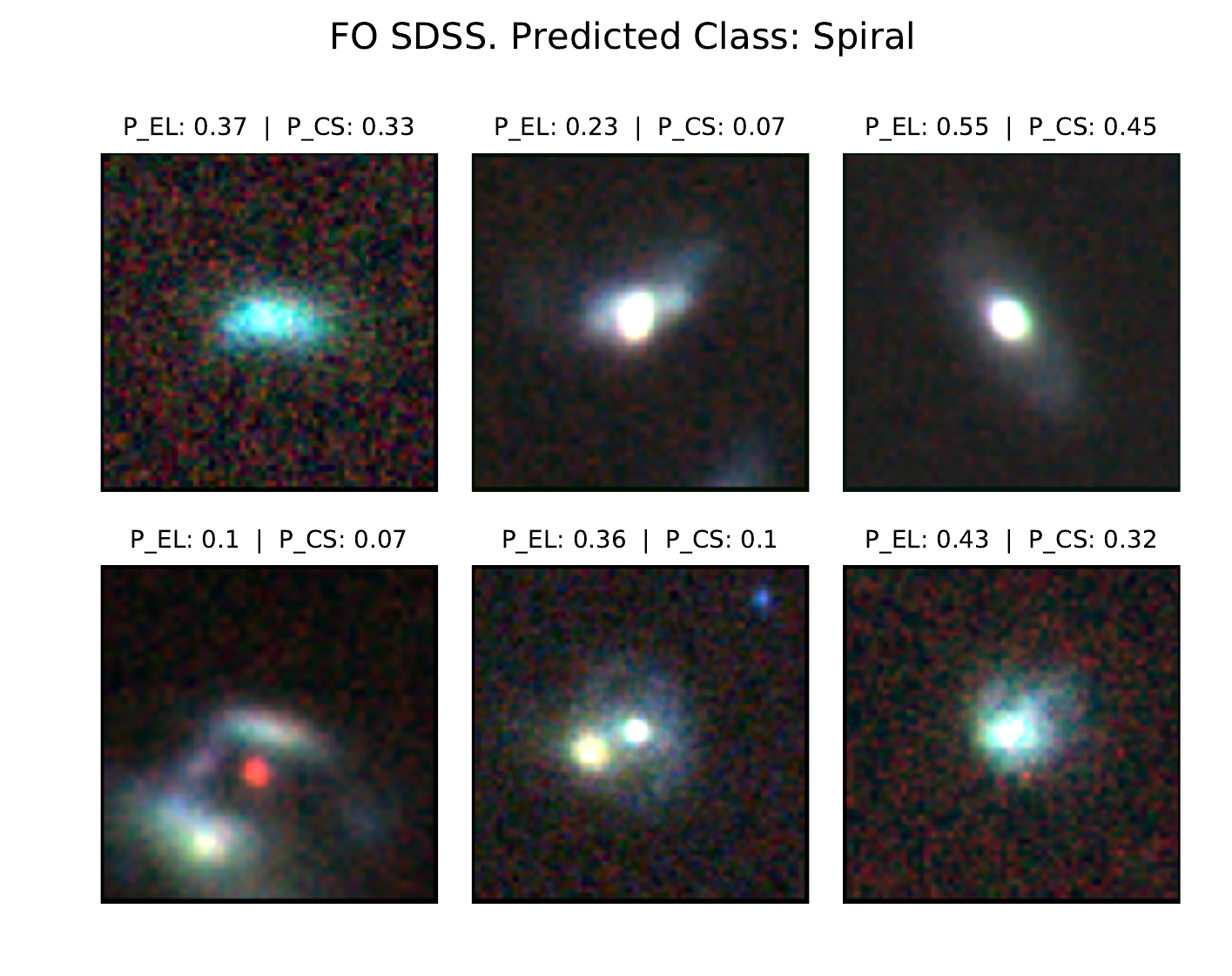}
}
\centerline{
\includegraphics[width=0.35\linewidth, height=0.25\linewidth]{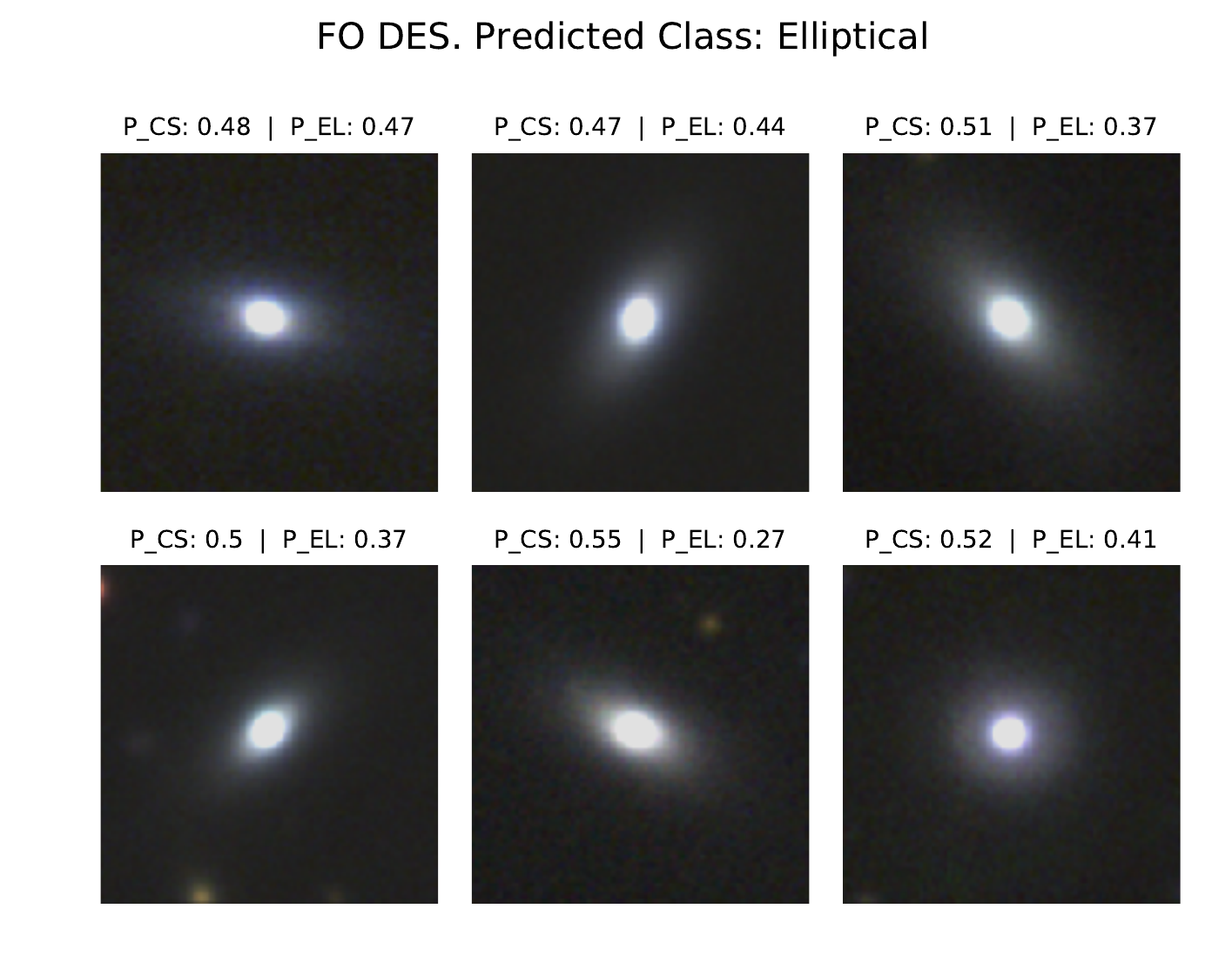}\hspace{-6mm}
\includegraphics[width=0.35\linewidth, height=0.25\linewidth]{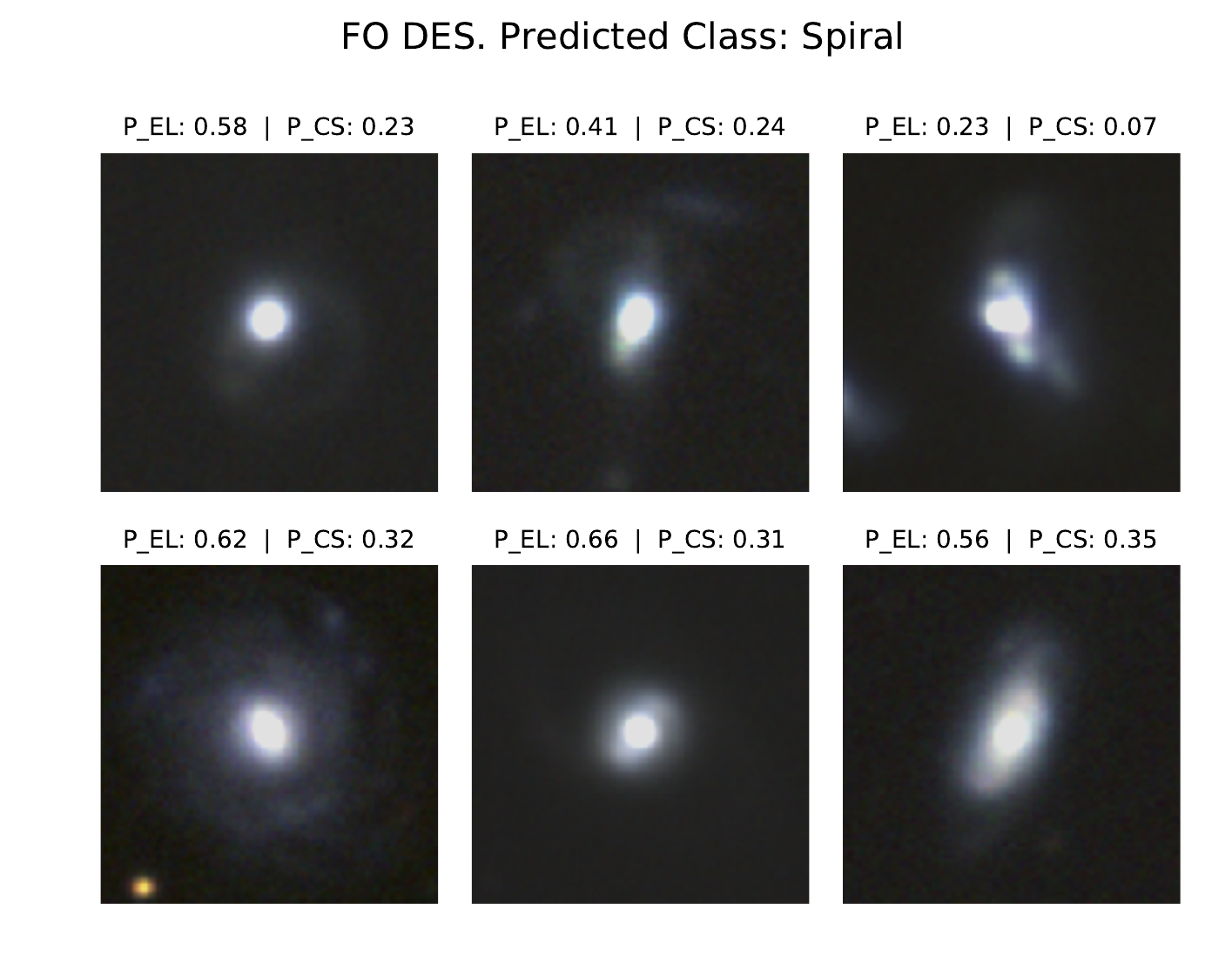}
}
\caption{A sample of misclassification on FO test sets. The debiased galaxy zoo probabilities used to produce the ground truth labels for each image are shown. Notice that the debiased galaxy zoo probabilities for each class are very low and close to each other, i.e., these samples represent low confidence ground truth labels.}
  \label{fig:misclassification}
\end{figure*}

\end{widetext}

%%%%%%%%%%%%%%%%%%%%%%%%%%%%%%%%%%%%%%%%%%%%%
%%%%%%%%%%%%%%%%%%%%%%%%%%%%%%%%%%%%%%%%%%%%%
\end{document}